\documentclass[12pt]{article}
\usepackage{latexsym,epsfig,graphicx,amsmath,amssymb,amscd,undertilde,multirow,chicago,psfrag,paralist,dsfont, url, xcolor,subfigure, array}
\newcommand{\thetavec}{{\boldsymbol{\theta}}}
\newcommand{\epsilonvec}{{\boldsymbol{\epsilon}}}

\newcommand{\Sigmavec}{{\boldsymbol{\Sigma}}}

\newcommand{\zerovec}{{\boldsymbol{0}}}

\newcommand{\lambdavec}{{\boldsymbol{\lambda}}}
\newcommand{\betavec}{{\boldsymbol{\beta}}}
\newcommand{\bvec}{{\boldsymbol{b}}}

\newcommand{\R}{\mathbb{R}}

\newcommand{\Var}{{\rm Var}}

\newcommand{\sigmahat}{\widehat{\sigma}}
\newcommand{\thetavechat}{\widehat{\thetavec}}

\newcommand{\APL}{{\rm APL}}
\newcommand{\PQL}{{\rm PQL}}
\newcommand{\tr}{{\rm tr}}

\newcommand{\Sigmatheta}{\Sigma_{\thetavec}}

\newcommand{\xvec}{\boldsymbol{x}}
\newcommand{\yvec}{\boldsymbol{y}}
\newcommand{\Lik}{\mathcal{L}}
\newcommand{\muvec}{\boldsymbol{\mu}}
\newcommand{\betavechat}{\widehat{\betavec}}
\newcommand{\bvechat}{\widehat{\bvec}}
\newcommand{\betavectilde}{\widetilde{\betavec}}
\newcommand{\thetavectilde}{\widetilde{\thetavec}}
\newcommand{\muvectilde}{\widetilde{\muvec}}
\newcommand{\bvectilde}{\widetilde{\bvec}}

\newcommand{\Corr}{{\rm Corr}}
\begin{document}
%%%%%%%%%%%%%%%%%%%%%%%%%%%%%%%%%%%%%%%%%%%%%%%%%%%%%%%%%%%%%%%%%%%%%%%%%%%%%%%%%%%%%%%%%%%%%%%%%%%%%%%%%%%%%%%%%

%%%%%%%%%%%%%%%%%%%%%%%%%%%%%%%%%%%%%%%%%%%%%%%%%%%%%%%%%%%%%%%%%%%%%%%%%%%%%%%%%%%%%%%%%%%%%%%%%%%%%%%%%%%%%%%%%
\title{Spatial Variable Selection and An Application to Virginia Lyme Disease Emergence}
%%%%%%%%%%%%%%%%%%%%%%%%%%%%%%%%%%%%%%%%%%%%%%%%%%%%%%%%%%%%%%%%%%%%%%%%%%%%%%%%%%%%%%%%%%%%%%%%%%%%%%%%%%%%%%%%%

%\iffalse
\author{Yimeng Xie$^1$, Li Xu$^1$, Jie Li$^1$, Xinwei Deng$^1$, Yili Hong$^1$, \\Korine Kolivras$^2$, and David N. Gaines$^3$\\[2ex]
{$^1$Department of Statistics, Virginia Tech, Blacksburg, VA 24061}\\
{$^2$Department of Geography, Virginia Tech, Blacksburg, VA, 24061}\\
{$^3$Virginia Department of Health, Richmond, VA, 23219}
}

%\fi

\date{}

%%%%%%%%%%%%%%%%%%%%%%%%%%%%%%%%%%%%%%%%%%%%%%%%%%%%%%%%%%%%%%%%%%%%%%%%%%%%%%%%%%%%%%%%%%%%%%%%%%%%%%%%%%%%%%%%%
\maketitle
%%%%%%%%%%%%%%%%%%%%%%%%%%%%%%%%%%%%%%%%%%%%%%%%%%%%%%%%%%%%%%%%%%%%%%%%%%%%%%%%%%%%%%%%%%%%%%%%%%%%%%%%%%%%%%%%%
\begin{abstract}
Lyme disease is an infectious disease that is caused by a bacterium called \emph{Borrelia burgdorferi} sensu stricto. In the United States, Lyme disease is one of the most common infectious diseases. The major endemic areas of the disease are New England, Mid-Atlantic, East-North Central, South Atlantic, and West North-Central. Virginia is on the front-line of the disease's diffusion from the northeast to the south. One of the research objectives for the infectious disease community is to identify environmental and economic variables that are associated with the emergence of Lyme disease. In this paper, we use a spatial Poisson regression model to link the spatial disease counts and environmental and economic variables, and develop a spatial variable selection procedure to effectively identify important factors by using an adaptive elastic net penalty. The proposed methods can automatically select important covariates, while adjusting for possible spatial correlations of disease counts. The performance of the proposed method is studied and compared with existing methods via a comprehensive simulation study. We apply the developed variable selection methods to the Virginia Lyme disease data and identify important variables that are new to the literature. Supplementary materials for this paper are available online.

\textbf{Key Words:} Gaussian Process; GLMM; Multicollinearity; Poisson Regression; Spatial Count Data; Spatial Correlation.
\end{abstract}
%%%%%%%%%%%%%%%%%%%%%%%%%%%%%%%%%%%%%%%%%%%%%%%%%%%%%%%%%%%%%%%%%%%%%%%%%%%%%%%%%%%%%%%%%%%%%%%%%%%%%%%%%%%%%%%%%
\newpage
%\tableofcontents
%\newpage

%%%%%%%%%%%%%%%%%%%%%%%%%%%%%%%%%%%%%%%%%%%%%%%%%%%%%%%%%%%%%%%%%%%%%%%%%%%%%%%%%%%%%%%%%%%%%%%%%%%%%%%%%%%%%%%%%
\section{Introduction} \label{sec:introduction}
%%%%%%%%%%%%%%%%%%%%%%%%%%%%%%%%%%%%%%%%%%%%%%%%%%%%%%%%%%%%%%%%%%%%%%%%%%%%%%%%%%%%%%%%%%%%%%%%%%%%%%%%%%%%%%%%%
Lyme disease is one of the most commonly reported vector-borne diseases in the United States. The disease was first identified in 1975 in the town of Old Lyme, Connecticut, and was therefore named Lyme disease. Lyme disease is caused by the bacterium \emph{Borrelia burgdorferi} sensu stricto. Through \emph{Ixodid} species tick bites, the bacterium is transmitted to humans. Early symptoms of Lyme disease includes skin rash, fever, headache, and fatigue. If the patients are not treated during the early stage of infection, severe and chronic symptoms can occur. Those chronic symptoms include arthritis in major joints, shooting pains, numbness in the hands or feet, and memory problems. Based on \citeN{Maes1998}, the estimated treatment costs for the disease were around \$2.5 billion over a five-year period, which is a significant public health burden. Thus, the study of Lyme disease emergence is of general interest in public health.

Although New England and other northeastern states are the initial endemic area for the disease, the endemic area has expanded over the last several decades, and Virginia is currently at its southward front line. Lyme disease spread from the northern part of Virginia to the southwestern part over the past decade, and the state also experienced an increasing number of cases. This makes Virginia an ideal state to study the mechanism behind the disease and discover crucial factors associated with emergence of Lyme disease. Because the transmission of Lyme disease involves tick bites, which are related to both  environmental and human factors, one of the important research questions in Lyme disease study is to identify possible environmental and demographic factors that can contribute to the emergence of the disease.

The main objective of this paper is to develop a method to identify a subset of the explanatory variables that are important for the case counts of Lyme disease based on Virginia data. As we can see from Lyme disease data (more details will be described in Section~\ref{sec:va.lyme.data}), there exist spatial correlations among disease counts, and there is also strong multicollinearity among explanatory variables. Variable selection while accounting for spatial dependence and multicollinearity is a challenging aspect. In the Lyme disease literature, the basic statistical models such as the Poisson regression are often used for modeling disease counts, without consideration of spatial correlation. When the goal is to identify important covariates, simple analyses such as bivariate analyses were used (e.g., \citeNP{Allan2003}, and \citeNP{Jackson2006}). The proposed methods in this paper will enable automatic variable selection while accounting for spatial correlation. Methodologically, the present work provides the Lyme disease research community more sophisticated analytic tools for statistical modeling and analysis. To the best of our knowledge, this paper is the first work that uses statewide Lyme disease data and covariates at census tract level to identify important environmental and human factors involved in the disease's spread.

Spatial data modeling, which has broad applications in ecology, epidemiology, agriculture, sociology, and other areas, has attracted great attention in recent years in the statistical literature.  For spatial data, correlations among observations in near locations are typically nonnegligible, and one way to model the spatial correlations among locations is through random effects. For example, \citeN{DiggleMoyeedTawn1998}, and \citeN{Zhang2002} employed generalized linear mixed models (GLMM) for spatial data with non-Gaussian outcomes. Various  approaches to estimate the parameters in GLMM have been developed. An overview on current methods can be found in \citeN{McCullochSearleNeuhaus2008}.

Regarding variable selections, a wide class of variable selection approaches have been developed via shrinkage methods.  The least absolute shrinkage and selection operator (LASSO) penalty is studied in \citeN{Tibshirani1996} to solve the regression type problem. It is shown that LASSO does parameter estimation and variable selection simultaneously due to the shrinkage property of the $L_{1}$ penalty. The ridge penalty introduced in \citeN{HorelKennard1970} always includes all the covariates. If there is a group of highly correlated covariates, the ridge penalty shrinks coefficients to each other but not to zero. Conversely, LASSO picks one covariate and assigns all weights to this covariate. In other words, ridge penalty tends to select the entire group, while LASSO tends to randomly pick only one covariate (\citeNP{Tibshirani1996}). \citeN{ZouHastie2005} proposed the elastic net penalty, which is a linear combination of the LASSO penalty and the ridge penalty. For a group of highly correlated covariates, the ridge and LASSO combination results in the trend of in and out together. Thus, the elastic net penalty has the property of automatic variable selection and continuous shrinkage.  LASSO does not have the oracle property and can be inconsistent unless certain conditions are satisfied. In light of these drawbacks of LASSO, the adaptive LASSO (\citeNP{Zou2006}) and adaptive elastic net (\citeNP{ZouZhang2009}) were developed. In addition, \citeN{FanLi2001} developed the smoothly clipped absolute deviation (SCAD) penalty. One may refer to \citeN{FanLv2010} for a comprehensive review of variable selection methods. In addition to the above work, Bayesian methods are also popular for variable selection. A review and comparison of Bayesian variable selection methods is available in \citeN{OharaSillanpaa2009}.

In terms of implementation, \shortciteN{Efronetal2004} proposed the least-angle regression (LARS) method to efficiently calculate the solution path of LASSO penalty in linear models. \citeN{ParkHastie2007} extended the concept of the LARS algorithm to generalized linear models (GLM). An algorithm named elastic net penalized least squares (LARS-EN) in \citeN{ZouHastie2005} is proposed for linear models with elastic net penalty. \citeN{SchelldorferMeierBuhlmann2014} developed ``GLMMLasso'' for high-dimensional GLMM with LASSO penalty and the corresponding R package is named ``glmmixedlasso'' (\citeNP{glmmixedlasso}). \citeN{GrollTutz2014} also considered this type of problem and a gradient descent algorithm is proposed to maximize the penalized log-likelihood function with implementation in an R package ``glmmLasso'' (\citeNP{glmmLasso}). A number of variable selection procedures for GLMM with longitudinal data settings are studied in \citeN{Yang2007} and \citeN{Cui2011}. Besides, \citeN{CaiDunson2006} proposed a fully Bayesian method to select fixed and random effects in the setting of GLMM. \shortciteN{Ibrahimetal2011} selected both fixed and random effects in a general class of mixed effects models using maximum penalized likelihood estimation along with the SCAD and the adaptive LASSO penalty functions. \citeN{YangZou2013} developed an generalized coordinate descent algorithm for computing the solution path of the hybrid Huberized support vector machine. \shortciteN{BoehmVocketal2015} developed spatial variable selection methods using a spatially varying coefficients model and applied them to study the acute health effects of fine particular matter components.

Despite the rich literature in spatial data analysis and various developments in variable selections, there is still a gap in performing variable selection for spatially correlated responses and correlated covariates. Although the focus of the paper is in Lyme disease applications, the developed spatial variable selection methods and comparisons also contribute to the general statistical literature. Extensive simulations show that the performance of the proposed methods perform better than existing methods for spatially correlated responses and correlated covariates. We also use bootstrap to quantify the uncertainties in parameter estimations. An R package is developed to implement the proposed methods.

The rest of this paper is organized as follows. Section~\ref{sec:va.lyme.data} introduces the Virginia Lyme disease data and the potential covariates for selection. Section~\ref{sec:data, model and likelihood} presents the spatial Poisson regression model with random effects and the computation of the likelihood function. Section~\ref{sec:estimation} presents two customized estimation procedures for estimating model parameters and develops a bootstrap algorithm for constructing confidence intervals. Section~\ref{sec:simulation study} conducts simulations to study the performance of the developed methods in variable selections and compares with existing methods. Section~\ref{sec:application to Lyme disease data} presents the data analysis for the Virginia Lyme data with interpretation and discussions. Section~\ref{sec:conclusion} contains conclusions and areas for future research.

%%%%%%%%%%%%%%%%%%%%%%%%%%%%%%%%%%%%%%%%%%%%%%%%%%%%%%%%%%%%%%%%%%%%%%%%%%%%%%%%%%%%%%%%%%%%%%%%%%%%%%%%%%%%%%%%%
\section{Virginia Lyme Disease Data}\label{sec:va.lyme.data}
%%%%%%%%%%%%%%%%%%%%%%%%%%%%%%%%%%%%%%%%%%%%%%%%%%%%%%%%%%%%%%%%%%%%%%%%%%%%%%%%%%%%%%%%%%%%%%%%%%%%%%%%%%%%%%%%%
The Lyme disease dataset for this paper contains case data from 2006 to 2011, and demographic data and land cover data in Virginia. Lyme disease case data were collected by the Virginia Department of Health (2006-\citeyearNP{VDH2006-2011}). The demographic data (e.g., population density, median income, and average age) are from the 2010 census (\citeNP{Almquist2010}). Land cover data were obtained from the Multi-Resolution Land Cover Consortium for 2006 (\shortciteNP{Fryetal2012}). A more detailed explanation of the data sources is available in \shortciteN{Seukepetal2015}.

\begin{figure}
\begin{center}
\begin{tabular}{cc}
\includegraphics[width=.45\textwidth]{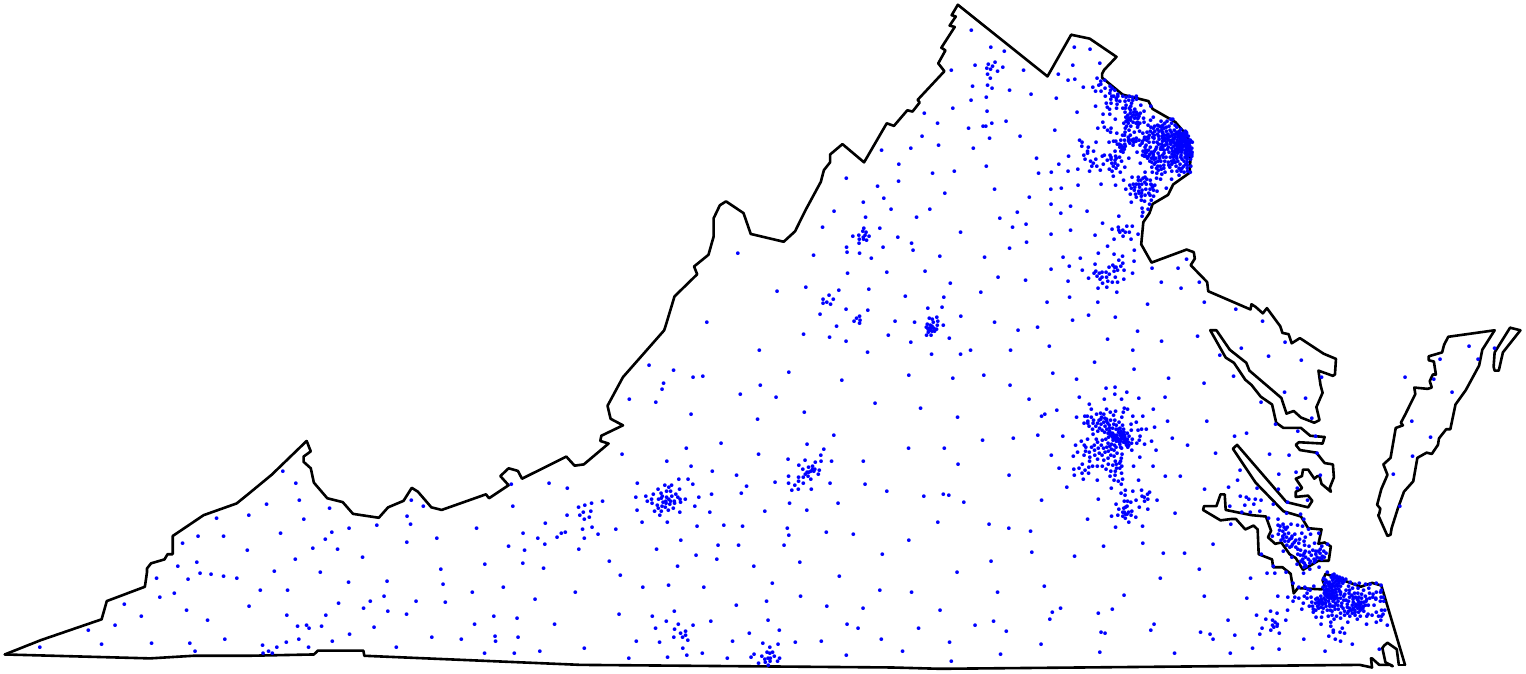}&
\includegraphics[width=.45\textwidth, angle=-8]{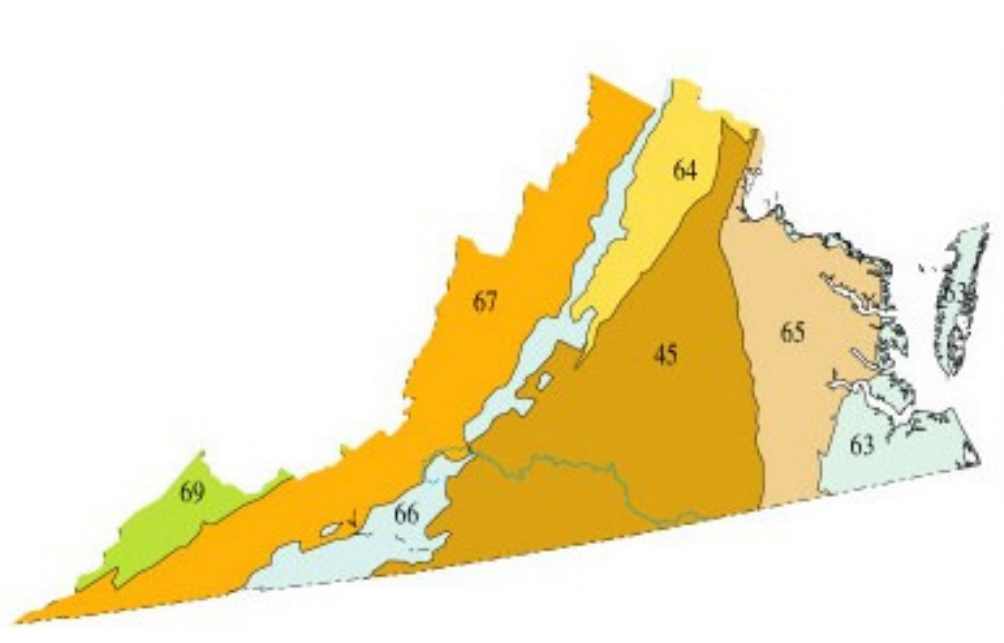}\\
(a) Census Centroids & (b) Ecoregions \\
\end{tabular}
\caption{(a) Illustrations of the Virginia study area and the locations of census centroids as indicated by dots, and (b) the level III ecoregions from Ecoregions of Virginia (2015). The Subregion~1 ($\textrm{Eco\_id}=1$) represents the southern/eastern subregion, which includes Piedmont (code 45), Middle Atlantic Coastal Plain (code 63), and Southeastern Plains (code 65). The Subregion~0 ($\textrm{Eco\_id}=0$) represents northern/western subregion, which includes Northern Piedmont (code 64), Blue Ridge (code 66), Ridge and Valley (code 67), and Central Appalachian (code 69).}\label{fig:ecoregions}
\end{center}
\end{figure}

\begin{figure}
\begin{center}
\begin{tabular}{c}
\includegraphics[width=0.7\textwidth]{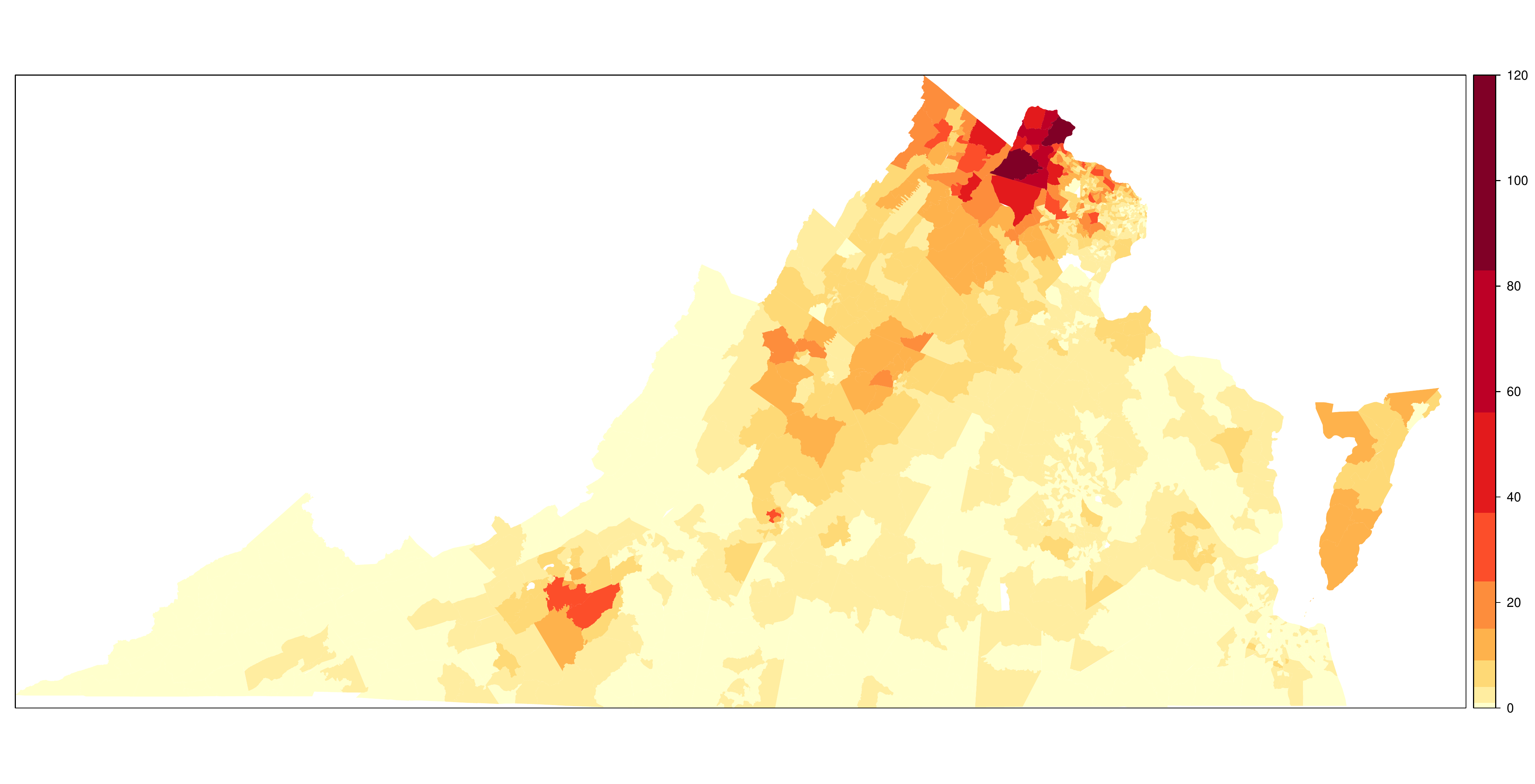}\\
(a) Case counts.\\
\includegraphics[width=0.7\textwidth]{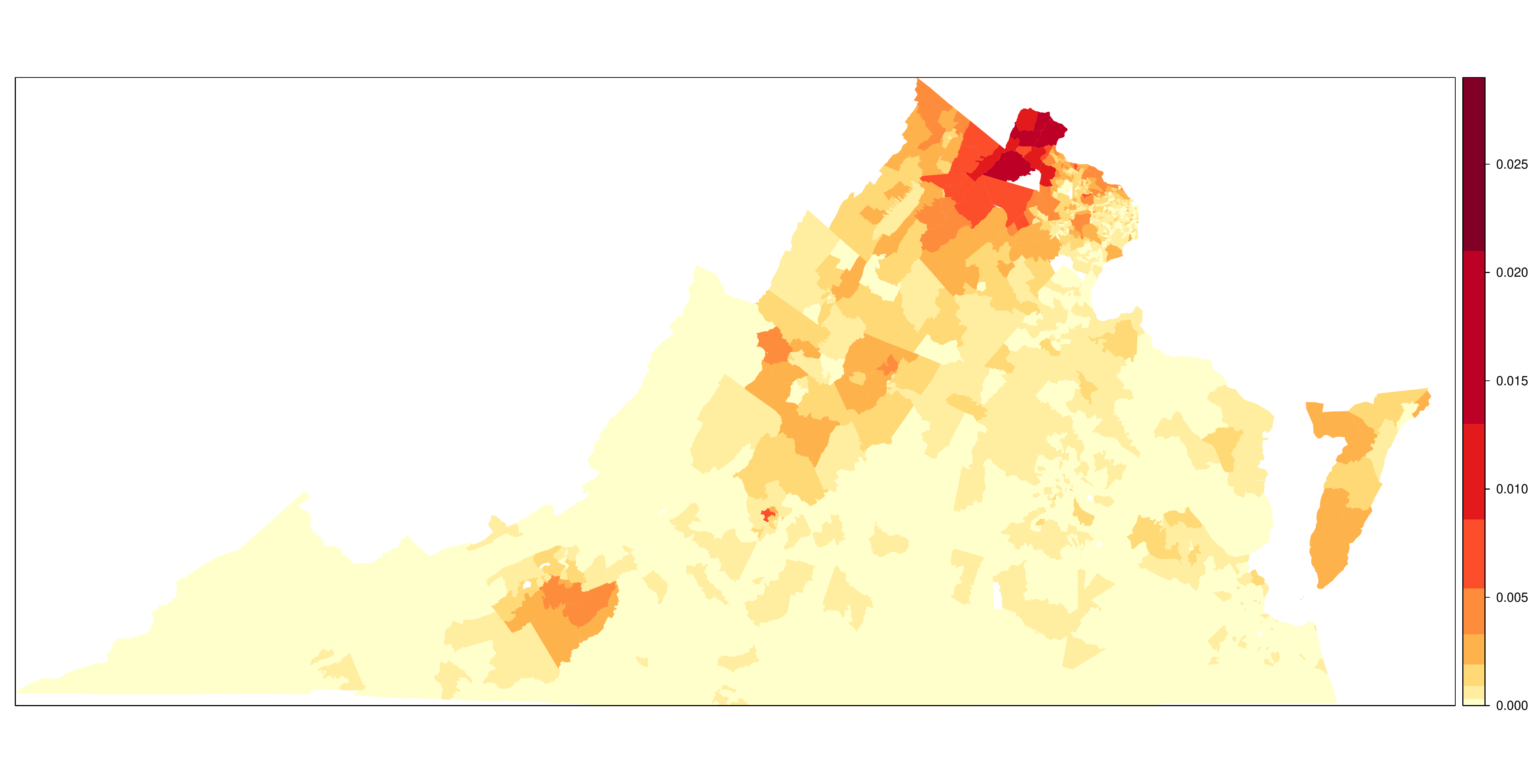}\\
(b) Incidence rates.
\end{tabular}
\end{center}
\caption{Number of cases and incidence rates of each census tract in Virginia for the five-year period (2006-2011). (a) Case counts. (b) Incidence rates.} \label{figure: lyme disease data}
\end{figure}

The Lyme disease cases were aggregated into census tracts for a couple of reasons: 1) the demographic information and land cover data are available for each census tract; and 2) the tract borders are based on certain features (e.g., rivers, roads) that are potential barriers to the movement of tick or other species involved in the Lyme disease transmission cycle (e.g., white-footed mice or deer). Figure~\ref{fig:ecoregions}(a) illustrates the Virginia study area and the locations of census centroids as indicated by dots. The response of interest is the summary of case counts from 2006 and 2011 in each census tract.  The total population counts in each census tract are included into the model as an offset term. Figure~\ref{figure: lyme disease data} shows the number of Lyme disease cases and incidence rates (i.e., the number of cases divided by the total population) for each census tract. A more detailed visualization of the Virginia Lyme disease data is available in \shortciteN{Lietal2014}.

Here, we discuss candidate covariates that may contribute to the case counts of Lyme disease.  The summary of the candidate covariates is available in Table~\ref{tab:list of covariates}. The dissimilarities in economic and demographic characteristics may affect the incidence of Lyme disease. To understand the transmission of Lyme disease, it is important to study the environment of ticks and Lyme disease reservoirs.

In past studies, white-footed mice or deer are shown to be very important hosts of ticks. Forested and herbaceous/scrub areas are ideal habitats for white-footed mice or deer.  According to \shortciteN{Seukepetal2015}, the herbaceous land type included scrub, herbaceous grasslands, cultivated agricultural lands, pasture, open impervious space, and emergent herbaceous wetlands. For example, scrub can provide tall enough vegetation to conceal deer and white-footed mice, and enough shade and humidity to allow black-legged ticks to survive the hot dry months of the summer. We consider four land cover types, which are the developed land, forest, herbaceous and water. The percentages of developed land, forest, and herbaceous within each tract are considered as covariates, with the percentage of water excluded because the percentages of those four land types add to 100\%.

\citeN{Allan2003} and \citeN{Jackson2006} studied the effect of forest fragmentation on Lyme disease and showed that the percent of forested areas and number of small forest fragments ($<$2 ha) within each polygon are associated with incidence rate of Lyme disease. In this study, we consider two types of forest fragmentation variables: percent of small forest fragments ($<$2 ha) and percent of perimeters of the small forest fragments ($<$2 ha) within each census tract.

The mixture of land cover types can also be an important factor for the disease cases. For example, the boundary between forest and residential areas raises the risk for the interaction between tick or disease reservoirs and humans, which may lead to an increase in the incidence rate. In this study, we consider three types of edges of land covers, which are the developed-forest edge, the forest-herbaceous edge, and the herbaceous-developed edge. For each edge, we consider two types of indices that characterize the mixture of land cover types:  Contrast Weighted Edge Density (CWED) and Total Edge Contrast Index (TECI), which represent two different algorithms for computing the mixture indices used in FRAGSTATS~4.1~(\citeNP{McGarigaletal2012}).

We also consider the type of ecoregion in our analysis. Based on the level III ecoregion map of Virginia~\citeyear{EcoregionVA}, Virginia can be divided into two major subregions, representing environmental and demographic differences. Figure~\ref{fig:ecoregions}(b) illustrates the level III Ecoregions of Virginia. Subregion~1 (i.e., the southern/eastern subregion) consists of Piedmont, Middle Atlantic Coastal Plain, and Southeastern Plains areas, while Subregion~0 (i.e., the northern/western subregion) includes the Northern Piedmont, Blue Ridge, Ridge and Valley and Central Appalachian areas. In addition, the population density, median age, and mean income in 2010 are also included as potential factors in the study.

We observe multi-collinearity among the covariates in the data using the pairwise correlations among all the covariates. The range of the absolute values of the pairwise correlations is from 0.002 to 0.986. From Supplementary Figure~1, we can also see that the correlations among the covariates can go quite high (i.e., above 0.8), which motivates us to use the elastic-net type penalty in variable selection.

\begin{table}
\begin{center}
\caption{Description of covariates in Lyme disease data.}\label{tab:list of covariates}
\vspace{1ex}
\begin{tabular}{l | l}\hline\hline
Variable           & Description \\\hline
Dvlpd\_NLCD06      & Percentage of developed land in each census tract\\
Forest\_NLCD06     & Percentage of forest in each census tract\\
Herbaceous\_NLCD06 & Percentage of herbaceous in each census tract\\
Tract\_Frag06      & Sum of area of forested fragments in each census tract\\[-.6ex]
                   & divided by the total area\\
FragPerim06        & Sum of forest fragment perimeters in each census tract\\[-.6ex]
                   & divided by the total area \\
CWED\_DF06         & CWED of developed-forest edge  \\
TECI\_DF06         & TECI of developed-forest edge \\
CWED\_FH06         & CWED of forest-herbaceous edge \\
TECI\_FH06         & TECI of forest-herbaceous edge \\
CWED\_HD06         & CWED of herbaceous-developed edge\\
TECI\_HD06         & TECI of herbaceous-developed edge \\
Pop\_den           & Tract population density in 2010\\
Median\_age        & Median age at each census tract in 2010 \\
Mean\_income       & Mean income (inflation adjusted) at each census tract\\[-.6ex]
                   & in 2010\\
Eco\_id            & $\textrm{Eco\_id}=1$ represents the Piedmont, Middle Atlantic \\[-.6ex]
                   & Coastal Plain, and Southeastern Plains areas;  $\textrm{Eco\_id}=0$ \\[-.6ex]
                   & represents the Northern Piedmont, Blue Ridge, Ridge \\[-.6ex]
                   & and Valley, and Central Appalachian areas\\\hline\hline
\end{tabular}
\end{center}
\end{table}

%%%%%%%%%%%%%%%%%%%%%%%%%%%%%%%%%%%%%%%%%%%%%%%%%%%%%%%%%%%%%%%%%%%%%%%%%%%%%%%%%%%%%%%%%%%%%%%%%%%%%%%%%%%%%%%%%
\section{The Statistical Model} \label{sec:data, model and likelihood}
%%%%%%%%%%%%%%%%%%%%%%%%%%%%%%%%%%%%%%%%%%%%%%%%%%%%%%%%%%%%%%%%%%%%%%%%%%%%%%%%%%%%%%%%%%%%%%%%%%%%%%%%%%%%%%%%%
Here we introduce some notations about the spatial count data and covariates. Let $n$ be the number of spatial locations, which are indexed by $i=1,\ldots, n$. Let $Y_i$ be the random variable for the count of disease cases at location $i$, which takes values in $\{0, 1, 2,\ldots\}$. The corresponding observation is denoted by $y_i$. The explanatory variables at location $i$ are denoted by $\xvec_i=(x_{i1},\ldots, x_{ij},\ldots,x_{ip})'$, where $p$ is the number of explanatory variables and $x_{ij}$ is the value of the $j$th covariate at location $i$. Let $\yvec=(y_{1}, y_{2}, \ldots, y_{n})'$ be the vector of the observations, and $X$ be an $n\times p$ matrix for the explanatory variables. That is $X=(\xvec_{1},\ldots, \xvec_{i},\ldots, \xvec_{n})'$. The population for location $i$ is denoted by $m_i$.

We use a spatial Poisson regression model with random effect $b_i$ to describe the spatial count data. That is,
\begin{align}\label{eqn:spatial Poisson regression model}
Y_i |b_{i} \sim \textrm{Poisson}(\mu_i),
\end{align}
where
$$\eta_i=\log(\mu_i)=\beta_0+x_{i1}\beta_1+\cdots+x_{ip}\beta_p+b_{i}+\log(m_i).$$
Here, $\mu_{i}$ is the conditional mean, $\eta_i$ is the log of $\mu_i$, $\beta_{j}$ is the regression coefficient of the corresponding covariate, and $\log(m_i)$ is the offset term corresponding to the population for location $i$. The random effect is $b_i$. Given $b_i$, the probability mass function (pmf) of $Y_i$ is $\exp{(-\mu_i)}\mu_{i}^{y_{i}}/y_{i}!$. The responses $Y_{i}$ are independent conditional on random effects $b_{i}$. Let $\muvec=(\mu_{1}, \ldots, \mu_{n})', \betavec=(\beta_{0}, \beta_{1},  \ldots, \beta_{p})'$, and $  \bvec=(b_1, b_{2}, \ldots, b_{n})'$.

The spatial correlations among locations are captured through the random effects $\bvec$. Following the spatial literature, we use the multivariate normal distribution to model the random effect $\bvec$. That is,
\begin{align}\label{eqn:bvec.dist}
\bvec \sim\textrm{N}(\zerovec,\Sigmatheta).
\end{align}
The variance-covariance matrix of $\bvec$ is $\Sigmatheta=\sigma^2\Omega$, and the $ij$th element of the $\Omega$ is $\rho(d_{ij};\thetavec)$. Here $\rho(\cdot)$ is a spatial correlation function and $\thetavec$ are parameters in $\Sigmatheta$. Note that $d_{ij}$ is the distance between two locations $i$ and $j$. In this paper, the exponential correlation function is used. That is, $\rho(d_{ij};\thetavec)=\exp(-d_{ij}/d)$ and $d>0$ is the scale parameter. In this case, $\thetavec=(\sigma^2, d)'$. The proposed method, however, can be extended to other spatial correlation functions such as the Gaussian, powered exponential, and Mat\'{e}rn correlation functions (e.g., \shortciteNP{Lietal2015}).

Note that we use a distance-based correlation structure in \eqref{eqn:bvec.dist}. In some disease mapping and ecology applications, the Gaussian Markov random field is also used for the correlation structure. However, our problem is special in the sense that we use census tracts as our study units. As illustrated in Figure~\ref{fig:ecoregions}(a), some census tracts in southwest Virginia are relatively large while other census tracts in the northern Virginia area (outside Washington DC) are relatively small. If one uses a Markov random field, it would ignore the differences in distance among different census tracts. In addition, the transmission of Lyme disease is related to distance. Based on those considerations, we use a distance-based correlation structure in this paper.

Based on the model specification in \eqref{eqn:spatial Poisson regression model} and \eqref{eqn:bvec.dist}, one can derive the likelihood function of unknown parameters $\betavec$ and $\thetavec$. Specifically, let $f(\yvec | \betavec, \bvec)$ be the pmf of $\yvec$ given $\bvec$, and $f(\bvec |\thetavec)$ be the probability density function (pdf) of $\bvec$. The likelihood function of $\{\betavec, \thetavec\}$ is
\begin{align} \label{eqn:likelihood function}
L(\betavec, \thetavec)
&=\int_{\R^{n}} f(\yvec | \betavec, \bvec) f(\bvec | \thetavec) \, d\bvec \nonumber \\
&=\int_{\R^{n}} \left[ \prod\limits_{i=1}^{n} \exp{(-\mu_i)} \dfrac{\mu_{i}^{y_{i}}}{y_{i}!} \right] \left[ (2\pi)^{-\frac{n}{2} } |\Sigmatheta|^{-\frac{1}{2}} \exp{\left(-\frac{1}{2} \bvec'\Sigmatheta^{-1}\bvec\right)} \right] \, d\bvec \nonumber \\
&= (2\pi)^{-\frac{n}{2} } |\Sigmatheta|^{-\frac{1}{2}} \int_{\R^{n}} \exp \left\{\sum\limits_{i=1}^{n} \left[ -\mu_{i}+y_{i}\eta_{i}-\log (y_{i}!) \right] - \frac{1}{2} \bvec'\Sigmatheta^{-1}\bvec \right\}  \, d\bvec,
\end{align}
where the integral of $\bvec$ is over the $n$-dimensional Euclidian space $\R^{n}$.

To perform variable selection, we add an adaptive elastic net (AEN) penalty term for fixed effects $\betavec$ to the log-likelihood function. That is, we consider the following penalized negative log-likelihood function
\begin{align} \label{eqn:objective function.exact}
\Lik(\betavec, \thetavec) = - \log[L(\betavec, \thetavec)] + P_{\lambdavec}(\betavec),
\end{align}
where
$$P_{\lambdavec}(\betavec)=\lambda_{1}\left[\lambda_{2}\sum_{j} \widehat{w}_{j} |\beta_j|+(1-\lambda_{2})\sum_{j}\beta_j^2\right]$$
is the AEN penalty. Here $\lambdavec=(\lambda_1, \lambda_2)'$ are regularization parameters. Note that $0\leqslant \lambda_2 \leqslant 1$, $\lambda_2=1$ is the case of LASSO penalty, and $\lambda_2=0$ is the case of ridge penalty. In addition, $\widehat{w}_{j}=|\betavechat_{\textrm{cnst}}|^{-r}$ is the adaptive weight with constant $r > 0$, and $\betavechat_{\textrm{cnst}}$ is an estimate of $\betavec$ that will be specified in Section~\ref{sec:Choice of Turning Parameters}.

Note that the likelihood function in (\ref{eqn:likelihood function}) contains intractable integrals over distribution of random effects. If the random effects are of low dimension, we may use Gaussian quadrature to do numerical integration. However, in the spatial Poisson regression model with random effects, the dimension of random effects is often the same as the number of observations. That is, the dimension of integrals is typically so large that the Gaussian quadrature or other low-dimensional methods may not work well. Because the computation of the exact likelihood is infeasible, if not impossible, approximate likelihood is often used in literature, by employing the Laplace approximation. In particular, the Laplace approximation (\citeNP{Laplace1986}) of multi-dimensional integrals over a multivariate function $\exp[h(\cdot)]$ is of the form
$$
\int_{\R^{n}} \exp[h(\bvec)] \, d\bvec \approx (2\pi)^{\frac{n}{2}} \bigg|-h''(\bvectilde)\bigg|^{-\frac{1}{2}} \exp[h(\bvectilde)],
$$
where $\bvectilde$ is the maximizer of function $h(\bvec)$, and $\big|-h''(\bvectilde)\big|$ is the determinant of the negative of the Hessian matrix of $h(\cdot)$. For the likelihood function in \eqref{eqn:likelihood function}, the corresponding $h(\cdot)$ function is
\begin{align}\label{eqn:hb.function}
h(\bvec)= \sum_{i=1}^{n} \left[ -\mu_{i}+y_{i}\eta_{i}-\log (y_{i}!) \right]-  \frac{1}{2}\bvec'\Sigma^{-1}\bvec.
\end{align}
Note that the maximizer $\bvectilde$ of the function $h(\cdot)$  in \eqref{eqn:hb.function} depends on parameters $\{\betavec, \thetavec\}$. Applying the Laplace approximation to likelihood function \eqref{eqn:likelihood function}, we obtain
\begin{align}\label{eqn:Laplace approximation to likelihood function}
\log\left[L(\betavec, \thetavec)\right]\approx l(\betavec, \thetavec),
\end{align}
where
$$l(\betavec, \thetavec)=-\frac{1}{2}\log\left(\big|\Sigmatheta W + I_{n}\big|\right)+\sum\limits_{i=1}^{n} \left[ -\mu_{i}+y_{i}\eta_{i}-\log (y_{i}!) \right] - \frac{1}{2} \bvectilde'\Sigmatheta^{-1}\bvectilde,$$
is the log of the approximate likelihood function, $W=\textrm{Diag}\{\muvec\}$, and $I_{n}$ is an $n \times n$ identity matrix.

Thus, we use the following approximate penalized log-likelihood (APL) function to approximate the objective function in \eqref{eqn:objective function.exact},
\begin{align} \label{eqn:objective function}
\Lik_{\APL}(\betavec, \thetavec) = - l(\betavec, \thetavec) + P_{\lambdavec}(\betavec).
\end{align}
The estimates of $\{\betavec, \thetavec\}$ can be obtained by finding the values of $\{\betavec, \thetavec\}$ that minimize the objective function in \eqref{eqn:objective function}. An alternative approach from \citeN{BreslowClayton1993} is to ignore the term $\big|\Sigmatheta W + I_{n}\big|$ in $l(\betavec, \thetavec)$, leading to the penalized quasi-likelihood (PQL) method. In particular, the PQL method aims to find the estimate of $\{\betavec, \thetavec\}$ by finding the values of $\{\betavec, \thetavec\}$ that minimize the following objective function,
\begin{align}\label{eqn:PQL.object.function}
\Lik_{\PQL}(\betavec, \thetavec) = - l_a(\betavec, \thetavec) + P_{\lambdavec}(\betavec),
\end{align}
where
\begin{align}\label{eqn:la.equation}
l_a(\betavec, \thetavec)=\sum\limits_{i=1}^{n} \left[ -\mu_{i}+y_{i}\eta_{i}-\log (y_{i}!) \right] - \frac{1}{2} \bvectilde'\Sigmatheta^{-1}\bvectilde.
\end{align}
Essentially, the PQL uses $l_a(\betavec, \thetavec)$ to approximate $l(\betavec, \thetavec)$. Under the consideration of computation, the PQL method is more efficient, with the tradeoff of ignoring the dependency of $\Sigmatheta$ and $W$ on $\{\betavec, \thetavec\}$ in the expression $\big|\Sigmatheta W + I_{n}\big|$.

%%%%%%%%%%%%%%%%%%%%%%%%%%%%%%%%%%%%%%%%%%%%%%%%%%%%%%%%%%%%%%%%%%%%%%%%%%%%%%%%%%%%%%%%%%%%%%%%%%%%%%%%%%%%%%%%%
\section{Parameter Estimation and Inference Procedures} \label{sec:estimation}
%%%%%%%%%%%%%%%%%%%%%%%%%%%%%%%%%%%%%%%%%%%%%%%%%%%%%%%%%%%%%%%%%%%%%%%%%%%%%%%%%%%%%%%%%%%%%%%%%%%%%%%%%%%%%%%%%
In this section, we develop computational methods to optimize $\Lik_{\APL}(\betavec, \thetavec)$ and $\Lik_{\PQL}(\betavec, \thetavec)$ in~\eqref{eqn:objective function} and \eqref{eqn:PQL.object.function}, respectively. The developed estimation procedures are iterative. The goal is to estimate the unknown parameters $\betavec$ and $\thetavec$. Both the APL and PQL methods share the following major steps:
\begin{itemize}
\item we first update $\bvectilde$ based on the current estimates of $\betavec$ and $\thetavec$,
\item then update $\betavec$ with penalty $P_{\lambdavec}(\betavec)$ to achieve variable selection, and
\item finally update $\thetavec$ based on the current estimates of $\betavec$ and $\bvectilde$.
\end{itemize}
The above three-step procedure will be conducted iteratively until convergence. The details for the selection of tuning parameters $\lambdavec$ are given in Section~\ref{sec:Choice of Turning Parameters}.

%%%%%%%%%%%%%%%%%%%%%%%%%%%%%%%%%%%%%%%%%%%%%%%%%%%%%%%%%%%%%%%%%%%%%%%%%%%%%%%%%%%%%%%%%%%%%%%%%%%%%%%%%%%%%%%%%
\subsection{The APL Method} \label{sec:Laplace Approximated Loglikelihood with Elastic Net Penalty}
%%%%%%%%%%%%%%%%%%%%%%%%%%%%%%%%%%%%%%%%%%%%%%%%%%%%%%%%%%%%%%%%%%%%%%%%%%%%%%%%%%%%%%%%%%%%%%%%%%%%%%%%%%%%%%%%%
For the APL method, the first step is to find $\bvectilde$ that maximizes $h(\bvec)$ in \eqref{eqn:hb.function}, given the current estimate of $\betavec$ and $\thetavec$. Regular optimization methods such as the Newton-Raphson method can be used here. The second step is to use the block coordinate gradient descent (BCGD) method in \citeN{TsengYun2009} to update $\betavec$ under penalty, given the current estimates of $\betavec$ and $\thetavec$. The solution of an AEN penalty problem can be solved by transforming it into a LASSO type of problem. Specifically, minimizing \eqref{eqn:objective function} with respect to $\betavec$ is equivalent to minimizing
$$
f(\betavec| \bvec ,\thetavec) + \lambda_1\lambda_2\sum_{j}\widehat{w}_{j}|\beta_j|,
$$
where
\begin{align}
f (\betavec| \bvec ,\thetavec) = \frac{1}{2} \log\left(\big|\Sigmatheta W+I_{n}\big|\right) - \sum\limits_{i=1}^{n} \left( -\mu_{i}+y_{i}\eta_i \right) + \frac{1}{2} \bvec'\Sigmatheta^{-1}\bvec + \lambda_1(1-\lambda_2)\sum_{j}\beta_j^2.
\end{align}
It is important to note that $f (\betavec| \bvec ,\thetavec)$ is a non-convex but differentiable function, and $\sum_{j}\widehat{w}_{j}|\beta_j|$ is a convex but non-differentiable function. To apply the BCGD algorithm, we update only one component of $\betavec$ at a time. For the $j$th component of $\betavec$, denoted as $\beta_j$, we first obtain $\bvectilde$ based on current estimates of $\betavec$ and $\thetavec$, which are denoted by $\betavectilde$ and $\thetavectilde$, respectively. Then we update the $j$th component by $\widetilde{\beta}_{j}+d_j$. Here,
$$
d_{j}=\textrm{median}\left\{\dfrac{\widehat{w}_{j}\lambda_1\lambda_2-f_{j}(\betavectilde| \bvectilde ,\thetavectilde)}{h_{jj}},\,\, -\widetilde{\beta}_{j},\,\, \dfrac{-\widehat{w}_{j}\lambda_1\lambda_2-f_{j}(\betavectilde| \bvectilde ,\thetavectilde)}{h_{jj}}\right\}.
$$
Also, $f_{j}(\betavec| \bvectilde ,\thetavectilde)$ is the $j$th component of the first derivative of $f(\betavec| \bvectilde ,\thetavectilde)$ with respect to $\betavec$ (i.e., $\partial f(\betavec| \bvectilde,\thetavectilde)/\partial\betavec$), and $h_{jj}$ is the $j$th diagonal element of $H$, where
\begin{align*}
\frac{\partial f(\betavec| \bvectilde,\thetavectilde)}{\partial\betavec}&=X'(\muvec
-\yvec)+2\lambda_1(1-\lambda_2)\betavec +\boldsymbol{c}, \\
H&=X'W X +2 \lambda_1(1-\lambda_2)I_{p}.
\end{align*}
The $j$th element of $\boldsymbol{c}$ is $(1/2)\tr\left\{(\Sigmatheta W+I_{n})^{-1}\Sigmatheta \partial W/\partial \beta_j\right\}$, and $I_{p}$ is a $p\times p$ identity matrix.

The last step of the iterative procedure is to update $\thetavec$. The estimate of $\thetavec$ is updated by minimizing \eqref{eqn:objective function} with current estimates $\betavectilde$ and $\bvectilde$. A description of the algorithm for the APL estimation procedure is as follows.
\\ \ \\
\textbf{Algorithm 1: APL with Adaptive Elastic Penalty (APL.AEN)}

For a collection of values of $(\lambda_1, \lambda_2)$:
\begin{enumerate}
\item Initialize $\betavec^{(0)}, \bvec^{(0)}$, and $\thetavec^{(0)}$.
\item For the $k$th iteration:
\begin{enumerate}[(i)]
\item To update the $j$th component of $\betavec$, one first finds $\bvectilde^{(k, j)}$ that maximizes $h(\bvec)$ with given $\betavectilde^{(k, j)}=(\widetilde{\beta}_{1}^{(k)}, \ldots, \widetilde{\beta}_{j-1}^{(k)}, \widetilde{\beta}_{j}^{(k-1)}, \ldots, \widetilde{\beta}_{p}^{(k-1)})$ and $\thetavectilde^{(k-1)}$.

\item Then update $\widetilde{\beta}_{j}^{(k)}=\widetilde{\beta}_{j}^{(k-1)}+d_j$.

\item Repeat (i) and (ii) for $j=1,\ldots, p$.

\item $\thetavectilde^{(k)}$ is obtained by minimizing \eqref{eqn:objective function} with $\betavectilde^{(k)}$.
\end{enumerate}
\item Repeat Step~2 until convergence. The final version of estimates $\betavectilde$, $\bvectilde$, and $\thetavectilde$ are denoted by $\betavechat, \widehat{\bvec},$ and $\thetavechat$, respectively.
\end{enumerate}

%%%%%%%%%%%%%%%%%%%%%%%%%%%%%%%%%%%%%%%%%%%%%%%%%%%%%%%%%%%%%%%%%%%%%%%%%%%%%%%%%%%%%%%%%%%%%%%%%%%%%%%%%%%%%%%%%
\subsection{The PQL Method} \label{sec:PQL with Elastic Net Penalty}
%%%%%%%%%%%%%%%%%%%%%%%%%%%%%%%%%%%%%%%%%%%%%%%%%%%%%%%%%%%%%%%%%%%%%%%%%%%%%%%%%%%%%%%%%%%%%%%%%%%%%%%%%%%%%%%%%
In this section we present how the parameters $\betavec$, $\thetavec$, and $\bvec$ are sequentially updated in the PQL method. We use the iterative algorithm in \citeN{BreslowClayton1993} to estimate $\{\betavec, \bvec\}$ from \eqref{eqn:hb.function}. In particular, \citeN{BreslowClayton1993} showed that the estimation of $\{\betavec, \bvec\}$ is equivalently to fit a linear mixed model (LMM) as follows:
\begin{align}\label{eqn:linear mixed model}
\yvec^{*}= X\betavec+\bvec+\epsilonvec, \textrm{ with } \bvec\sim \mathrm{N}(\zerovec, \Sigmatheta), \textrm{ and } \epsilonvec \sim \mathrm{N}(\zerovec, W^{-1}).
\end{align}
Here $\yvec^{*}=(y_{1}^{*}, \ldots, y_{n}^{*})$ is the working response vector with $y_{i}^{*}=\xvec_{i}'\betavec+b_{i}+{(y_i-\mu_i)}/{\mu_i}$. Based on the model formulation in \eqref{eqn:linear mixed model}, we obtain $\Var(\yvec^{*})=V=W^{-1}+\Sigmatheta$. We update $\betavectilde$ and $\bvectilde$ in the following formulas iteratively until convergence. In particular,
\begin{gather*}
\betavectilde=\left(X'V^{-1}X\right)^{-1}X'V^{-1}\yvec^{*},\quad \textrm{and}\quad \bvectilde=\Sigmatheta V^{-1} \left(\yvec^{*}-X\betavectilde\right).
\end{gather*}

The second step is to update $\betavec$ under penalty. To efficiently obtain one step update of $\betavec$ under the penalty, we use a quadratic approximation to $l_{a}(\betavec| \bvectilde, \thetavectilde)$, which is the $l_a(\betavec, \thetavec)$ in \eqref{eqn:la.equation} but given $\thetavectilde$ and $\bvectilde$. Note that $ l_{a}(\betavec| \bvectilde, \thetavectilde)$ is a concave function of $\betavec$ given $\thetavectilde$ and $\bvectilde$, and $\betavectilde$ is the maximizer of $l_{a}(\betavec| \bvectilde, \thetavectilde)$. We form a quadratic approximation to $ l_{a}(\betavec| \bvectilde, \thetavectilde)$ around $\betavectilde$ to speed up the updating of $\betavec$. \citeN{FriedmanHastieTibshirani2010} used a similar method for generalized linear models. In particular, we have
\begin{align*}
l_{a}(\betavec|\thetavectilde, \bvectilde) \approx -\dfrac{1}{2} \sum\limits_{i=1}^{n} \mu_{i}(z_{i}-\xvec_i'\betavec)^2,
\end{align*}
where $z_{i}=\xvec_i'\betavectilde-1+y_{i}/\mu_{i}$. The details regrading the derivation of the quadratic approximation is given in Appendix \ref{sec:quadratic approximation}. Incorporating the elastic net penalty function, we obtain
\begin{align} \label{eqn:PQL and quadratic approximation with penalty}
\Lik_{\textrm{PQL}}^{Q}(\betavec|\thetavectilde, \bvectilde) =  \dfrac{1}{2} \sum\limits_{i=1}^{n} \mu_{i}(z_{i}-\xvec_i'\betavec)^2 + P_{\lambdavec}(\betavec).
\end{align}
The minimizer $\betavectilde_{Q}$ of (\ref{eqn:PQL and quadratic approximation with penalty}) can be achieved as a penalized weighted least squares problem by using the R package ``\textit{glmnet}'' (\citeNP{FriedmanHastieTibshirani2010}).

It is interesting to point out that estimating $\betavec$ via optimizing (\ref{eqn:PQL and quadratic approximation with penalty}) is equivalent to consider the penalized log-likelihood of the linear mixed model (\ref{eqn:linear mixed model}). That is,
\begin{align} \label{eqn:double penalized loglikelihood function of LMM}
 -\frac{1}{2} (\yvec^{*}-X\betavec-\bvec)' W (\yvec^{*}-X\betavec-\bvec) - \frac{1}{2} \bvec'\Sigmatheta^{-1}\bvec- P_{\lambdavec}(\betavec).
\end{align}
To see the connection between (\ref{eqn:PQL and quadratic approximation with penalty}) and (\ref{eqn:double penalized loglikelihood function of LMM}), we estimate $\bvectilde$ first and let $\yvec^{**}=\yvec^{*}-\bvectilde$. Then again we have a weighted linear regression with elastic net penalty problem. That is we want to minimize
$$
\frac{1}{2}\sum\limits_{i=1}^{n}\mu_i(y_i^{**}-\xvec_{i}'\betavec)^2 + P_{\lambdavec}(\betavec),
$$
which is equivalent to minimizing (\ref{eqn:PQL and quadratic approximation with penalty}). That is, we can transform the GLMM with penalty problem into an LMM with penalty problem.

The last step of the iterative procedure is to update $\thetavec$. We update $\thetavec$ by using the restricted maximum likelihood (REML) method. The calculation of the REML function involves the elastic net penalty function $P_{\lambdavec}(\betavec)$, which has the singularity at the origin. Therefore, we consider an approximation of the penalty function $P_{\lambdavec}(\betavec)$. Based on \citeN{FanLi2001}, the penalty function $P_{\lambdavec}(\betavec)$ can be approximated by
$$
P_{\lambdavec}(\betavec) \approx \frac{1}{2}\betavectilde_{\lambdavec}'\Sigmavec_{\lambdavec}(\betavectilde)\betavec_{\lambdavec},
$$
where $\betavectilde_{\lambdavec}$ only contains nonzero elements $\widetilde{\beta}_{1},\ldots, \widetilde{\beta}_{m}$ of $\betavectilde$, and $$\Sigmavec_{\lambdavec}(\betavectilde)=\textrm{Diag}\left\{\frac{P_{\lambdavec,\, \widetilde{\beta}_{1}}(|\widetilde{\beta}_{1}|)}{|\widetilde{\beta}_{1}|}, \ldots, \frac{P_{\lambdavec,\, \widetilde{\beta}_{m}}(|\widetilde{\beta}_{m}|)}{|\widetilde{\beta}_{m}|} \right\}.$$
Here $P_{\lambdavec,\, \widetilde{\beta}_{j}}(|\widetilde{\beta}_{j}|)$ is the first partial derivative with respect to $\widetilde{\beta}_{j}$. Also, define $X_{\lambdavec}$ to be the matrix corresponding to the nonzero elements of $\betavectilde$. \citeN{Cui2011} showed that the approximate REML estimator for $\thetavec$ can be calculated by maximizing
\begin{align} \label{eqn:Cui(2011) REML loglikelihood for covariance parameters}
-\frac{1}{2}\log |V|-\frac{1}{2} \log |X_{\lambdavec}'V^{-1}X_{\lambdavec}+\Sigmavec_{\lambdavec}(\betavectilde)|-\frac{1}{2} \left(\yvec^{*}-X_{\lambdavec}\betavectilde_{\lambdavec}\right)'V^{-1}\left(\yvec^{*}-X_{\lambdavec}\betavectilde_{\lambdavec}\right).
\end{align}
Note that the term $\Sigmavec_{\lambdavec}(\betavectilde)$ in $\log |X_{\lambdavec}'V^{-1}X_{\lambdavec}+\Sigmavec_{\lambdavec}(\betavectilde)|$ is for adjustment of the penalty function of $\betavec$. Note that in alternative of maximizing the function in \eqref{eqn:Cui(2011) REML loglikelihood for covariance parameters}, expectation maximization (EM) type estimators can also be used (e.g., \citeNP{FahrmeirTutz2001}, and \citeNP{GrollTutz2014}). The estimation procedure is summarized in the following algorithm.

\vspace{2ex}
\noindent\textbf{Algorithm 2: PQL with Adaptive Elastic Net Penalty (PQL.AEN)}\\
For a collection of values of $(\lambda_1, \lambda_2)$:
\begin{enumerate}
\item Initialize $\betavec^{(0)}, \bvec^{(0)}$, and $\thetavec^{(0)}$.
\item For the $k$th iteration:
\begin{enumerate}[(i)]
\item Find $\{\betavec, \bvec\}$ that maximize \eqref{eqn:hb.function}. Specifically, define the working response as $$\yvec^{*(k)}=\xvec_{i}'\betavectilde^{(k-1)}+\bvectilde^{(k-1)}+{(\yvec-\muvectilde^{(k-1)})}/{\muvectilde^{(k-1)}},$$  and update $\betavectilde$ and $\bvectilde$ iteratively until converge. The estimates obtained are denoted as $\betavectilde^{(k)}, \bvectilde^{(k)}$.
\item Given the current estimates $\betavectilde^{(k)}, \bvectilde^{(k)},$ and  $\thetavectilde^{(k-1)}$, solve
\begin{align}
\Lik_{\textrm{PQL}}^{Q}(\betavec|\thetavectilde^{(k-1)}, \bvectilde^{(k)}) = \dfrac{1}{2} \sum\limits_{i=1}^{n} \mu_{i}(z_{i}-\xvec_i'\betavec)^2 + P_{\lambdavec}(\betavec),
\end{align}
where $z_{i}$ and $\mu_{i}$ are evaluated at $\betavectilde^{(k)}, \bvectilde^{(k)},$ and $\thetavectilde^{(k-1)}$. The estimate obtained is denoted by $\betavectilde_{Q}^{(k)}$.
\item Obtain the estimates of covariance parameters $\thetavectilde^{(k)}$ by maximizing (\ref{eqn:Cui(2011) REML loglikelihood for covariance parameters}) with $\betavectilde_{Q}^{(k)}$ and $\bvectilde^{(k)}$.
\end{enumerate}
\item Repeat Step 2 until convergence. The final version of estimates $\betavectilde_{Q}$, $\bvectilde$, and $\thetavectilde$ are denoted by $\betavechat, \widehat{\bvec},$ and $\thetavechat$, respectively.
\end{enumerate}

Both \textbf{Algorithms~1} and~\textbf{2} are implemented in R~\citeyear{R2016} via an R package ``SpatialVS''~\citeyear{SpatialVS-Rpackage} and a data package ``VALymeData''~\citeyear{SpatialVS-Datapackage}. The Virginia Lyme disease data and the R code for simulation and analysis are also available via the online supplementary materials.

%%%%%%%%%%%%%%%%%%%%%%%%%%%%%%%%%%%%%%%%%%%%%%%%%%%%%%%%%%%%%%%%%%%%%%%%%%%%%%%%%%%%%%%%%%%%%%%%%%%%%%%%%%%%%%%%%
\subsection{Specification of Adaptive Weights and Selection of Tuning Parameters} \label{sec:Choice of Turning Parameters}
%%%%%%%%%%%%%%%%%%%%%%%%%%%%%%%%%%%%%%%%%%%%%%%%%%%%%%%%%%%%%%%%%%%%%%%%%%%%%%%%%%%%%%%%%%%%%%%%%%%%%%%%%%%%%%%%%
We need to specify the adaptive weights $\widehat{w}_{j}=|\betavechat_{\textrm{cnst}}|^{-r}$ for the AEN penalty in \eqref{eqn:objective function.exact}. Following~\citeN{ZouZhang2009}, we specify $\betavechat_{\textrm{cnst}}$ to be the estimates under the elastic net penalty. For those elements of $\betavechat_{\textrm{cnst}}$ that are set to zero by the elastic net penalty, we set them to be $1/n$ as in~\citeN{ZouZhang2009}. We use $r=1$ in the simulation study and data analysis.

Regarding tuning parameters, popular methods of choosing the tuning parameters $\lambdavec=(\lambda_1, \lambda_2)'$ include cross-validation and criterion-based approaches. In this paper, we use the Bayesian Information Criterion (BIC) to select the tuning parameter. The calculation of exact log-likelihood for GLMM is complicated. Thus the Laplace approximated log-likelihood is used. For notation simplicity, we also use $\betavechat,  \bvechat,$ and $\thetavechat$ to represent estimates obtained from penalized approximate likelihood. In particular, the BIC is defined by $-2 l(\betavechat, \thetavechat) + \log(n) df$,
where $df$ is number of nonzero parameters in $\betavechat$ plus the number of parameters in $\thetavechat$. The values of the tuning parameters $\lambdavec=(\lambda_1, \lambda_2)'$ are chosen to minimize the BIC.

Here we provide a brief discussion on the degrees of freedom (df) of the model. We use the number of nonzero parameters in $\betavechat$ plus the number of parameters in $\thetavechat$ as the effective df, following \shortciteN{SchelldorferMeierBuhlmann2014} and \citeN{GrollTutz2014}. The ``real'' complexity in GLMM is an open issue of current research. In particular, not only the number of random effects variance-covariance parameters but also their size has an influence on the model's complexity. For example, if a random effect has a large variance, the corresponding random intercept or slope estimates are much larger and, the model tends to be more complex. Alternatively, the ``glmmLasso'' in \citeN{glmmLasso} allows to use the trace of the corresponding approximate hat matrix as model complexity.

%%%%%%%%%%%%%%%%%%%%%%%%%%%%%%%%%%%%%%%%%%%%%%%%%%%%%%%%%%%%%%%%%%%%%%%%%%%%%%%%%%%%%%%%%%%%%%%%%%%%%%%%%%%%%%%%%
\subsection{Confidence Interval Procedures}
%%%%%%%%%%%%%%%%%%%%%%%%%%%%%%%%%%%%%%%%%%%%%%%%%%%%%%%%%%%%%%%%%%%%%%%%%%%%%%%%%%%%%%%%%%%%%%%%%%%%%%%%%%%%%%%%%
In this section, we first review several existing methods for statistical inference of penalized models and then suggest to use parametric bootstrap to obtain confidence intervals~(CIs) for parameters in the spatial model in this paper. \citeN{FanLi2001} derived a sandwich-type standard error formula for nonzero components of the LASSO estimator. \citeN{Zou2006} used a similar approach to derive the standard error formula for the nonzero components of the adaptive Lasso estimator. The sandwich-type estimator, however, can not provide uncertainty quantification for those zero components of the estimator. \shortciteN{berk2013valid} proposed a framework for valid post-selection inference by using simultaneous inference. \citeN{Bachocetal2016} proposed a general method to construct asymptotically uniformly valid CIs post-model-selection using the principles in \shortciteN{berk2013valid}. \shortciteN{lee2016exact} developed a general approach for valid inference after model selection by characterizing the distribution of the post selection-estimator conditional on the selection event. \shortciteN{lu2017confidence} investigated the CI problem from a different point of view, and they used stochastic variational inequality techniques in optimization to derive CIs for the LASSO estimator. Overall, the current methodological developments are mostly made for LASSO type estimators.

\citeN{ning2017general} developed general theory for statistical inference for generic penalized M-estimator using the idea of decorrelated score function. Their approach is quite general and it can be applied to a variety of models such as linear models, generalized linear models, and survival models. Their approach can not be directly applied to our setting because all observations are correlated under the spatial model.

\citeN{chatterjee2011bootstrapping} showed that bootstrap methods are valid for the adaptive LASSO estimator due to its oracle property. For the AEN penalty, \citeN{ZouZhang2009} showed that it also has oracle property under the setting of linear models. Although our setting is different, based on current methodological developments, a practical approach for constructing CIs for our model is the parametric bootstrap. One advantage of using the parametric bootstrap is that it can easily keep the spatial correlation in the bootstrapped samples. Thus, we use the parametric bootstrap to construct CIs in this paper. The detailed algorithm for the parametric bootstrap is available in Supplementary Section~2.

%%%%%%%%%%%%%%%%%%%%%%%%%%%%%%%%%%%%%%%%%%%%%%%%%%%%%%%%%%%%%%%%%%%%%%%%%%%%%%%%%%%%%%%%%%%%%%%%%%%%%%%%%%%%%%%%%
\section{Simulation Studies} \label{sec:simulation study}
%%%%%%%%%%%%%%%%%%%%%%%%%%%%%%%%%%%%%%%%%%%%%%%%%%%%%%%%%%%%%%%%%%%%%%%%%%%%%%%%%%%%%%%%%%%%%%%%%%%%%%%%%%%%%%%%%
In this section, we evaluate the performance of the APL and PQL methods proposed in Sections~\ref{sec:Laplace Approximated Loglikelihood with Elastic Net Penalty} and \ref{sec:PQL with Elastic Net Penalty}, and compare with existing methods through simulations.

%%%%%%%%%%%%%%%%%%%%%%%%%%%%%%%%%%%%%%%%%%%%%%%%%%%%%%%%%%%%%%%%%%%%%%%%%%%%%%%%%%%%%%%%%%%%%%%%%%%%%%%%%%%%%%%%%
\subsection{Simulation Setting} \label{sec:simulation settings}
%%%%%%%%%%%%%%%%%%%%%%%%%%%%%%%%%%%%%%%%%%%%%%%%%%%%%%%%%%%%%%%%%%%%%%%%%%%%%%%%%%%%%%%%%%%%%%%%%%%%%%%%%%%%%%%%%
In the simulation study, we consider the following model:
\begin{gather*}
y_{i} | b_{i} \sim \textrm{Poisson} \left[\exp(\xvec_{i}'\betavec + b_{i}) \right],
\end{gather*}
where $\xvec_i$ is the vector of covariates and $\betavec$ collects the corresponding coefficients. Here, the distribution of the random effect $\bvec$ is the same as in \eqref{eqn:bvec.dist}. That is the covariance has the form $(\Sigmatheta)_{ij}=\sigma^{2}\exp(d_{ij}/d)$, where $d>0$ is the scale parameter. Each dataset consists of $n=225$ equal spaced data points that are simulated on a $[1, 10] \times [1, 10]$ regular grid. The distance between point $i$ and $j$ is denoted by $d_{ij}$. The $\xvec_{i}$ are simulated from multivariate normal distribution with mean $0$ and variance $0.5$.

We consider the following three settings of $\betavec$ and $\thetavec$ to represent different degrees of covariate effects and the number of active (i.e., nonzero-effect) covariates:

\begin{inparaenum}[(i)]
\item $\betavec=(-0.5, 0.75, 1, -0.75, -1, \zerovec_{10})'$,

\item $\betavec=(0.2, 0.3, 0.4, 0.5, 0.7, 0.8, -0.1, -0.6, -0.9, -1, \zerovec_{10})'$,

\item $\betavec=(-0.5, 0.75, 1, -0.75, -1, \zerovec_{20})'$.
\end{inparaenum}
\\
Here $\zerovec_{n}$ is a vector of zeros with length $n$, and let $p$ represent the length of $\betavec$. The value of $\thetavec=(\sigma^{2}, d)'$ is specified to be $(0.1, 5)', (0.5, 5)'$ or $(0.1, 10)'$.

For the model matrix, we consider the following five cases to represent various types of collinearity among covariates. The main motivation is to explore different kind of correlation structures to see if there are any effects on the variable selection.

\begin{inparaenum}
\item All covariates are independent.

\item $\Corr(X_{k}, X_{l})=\omega^{|k-l|}, k=1,\dots, 5, l=1,\dots, 5$ with $\omega=0.8$; the other covariates are independent. In this case, the first five covariates are correlated with exponential decay.

\item $\Corr(X_{k}, X_{l})=\omega^{|k-l|}, k=1,2,3, l=1,2,3$ with $\omega=0.8$ and $\Corr(X_{4}, X_{5})=0.8$; the other covariates are independent. In this case, we consider different degree of correlations among covariates. We impose a strong correlation between the first two covariates. Another three variables are correlated with exponential decay correlations, and the rest variables are uncorrelated.

\item $\Corr(X_{k}, X_{l})=\omega^{|k-l|}, k=1,2,3, l=1,2,3$ with $\omega=0.8$ and $\Corr(X_{4}, X_{5})=0.5$; the other covariates are independent. This case is similar to Case~3 but with two moderately correlated covariates.

\item $\Corr(X_{k}, X_{l})=\omega^{|k-l|}, k=1,\dots, 5, l=1,\dots, 5$ with $\omega=0.8$; $\Corr(X_{k}, X_{l})=\omega^{|k-l|}, k=p-4,\dots, p, l=p-4,\dots, p$ with $\omega=0.8$; the other covariates are independent. While Cases~2-4 only consider nonzero-effect covariates to be correlated, Case~5 extends to the scenario that zero-effect covariates can also be correlated.

\end{inparaenum}

For each case, we simulate 300 datasets and the covariates are all centered and standardized. For simplicity, we assume there is no intercept term in the model. For each simulated dataset, we apply the methods described in Sections~\ref{sec:Laplace Approximated Loglikelihood with Elastic Net Penalty} (APL.AEN) and \ref{sec:PQL with Elastic Net Penalty} (PQL.AEN) to obtain estimates of parameters and do variable selection. We also fit the case of $(\Sigmatheta)_{ij}=\sigma^{2}$, under which case the spatial correlation induced by the distance is ignored.

We consider the following performance measures for variable selection accuracy: (a) \textit{aver.size}: average model size; (b) \textit{corr.coef}: average number of coefficients set to 0 correctly; (c) \textit{mis.coef}: average number of coefficients set to 0 incorrectly.

%%%%%%%%%%%%%%%%%%%%%%%%%%%%%%%%%%%%%%%%%%%%%%%%%%%%%%%%%%%%%%%%%%%%%%%%%%%%%%%%%%%%%%%%%%%%%%%%%%%%%%%%%%%%%%%%%
\subsection{Results and Discussions} \label{sec:simulation results and discussions}
%%%%%%%%%%%%%%%%%%%%%%%%%%%%%%%%%%%%%%%%%%%%%%%%%%%%%%%%%%%%%%%%%%%%%%%%%%%%%%%%%%%%%%%%%%%%%%%%%%%%%%%%%%%%%%%%%
Table~\ref{tab:Model selection results based on simulated samples.} reports the \textit{aver.size}, \textit{corr.coef} and \textit{mis.coef} for the setting of $\betavec=(-0.5, 0.75, 1, -0.75, \\ -1,  \zerovec_{10})'$, and $\thetavec=(0.1, 5)'$.  There is no big difference among the five cases of model matrices, which suggests that the AEN penalty performs well for correlated covariates. The APL or PQL methods yield similar results. Considering spatial correlation yields slightly better results than ignoring spatial correlation.

Table~\ref{tab:Model selection results based on simulated samples, increase sigma_sq.} summarizes the results of considering $\betavec=(-0.5, 0.75, 1, -0.75, -1, \zerovec_{10})'$, and $\thetavec=(0.5, 5)'$. The APL method provides reasonably good results, while the PQL method gives slightly worse results because the PQL method tends to have larger active sets. Comparing to the results in Table~\ref{tab:Model selection results based on simulated samples.}, for the PQL method, the average number of coefficients that is set to 0 correctly is lower and the average model size is larger. If $\sigma^{2}$ increases, which means the random effects account for greater proportion of variation in the dependent variable, the PQL method tends to include more irrelevant covariates, while the performance of the APL method is less affected. Table~\ref{tab:Model selection results based on simulated samples, increased d.} shows the results of increasing $d$ (i.e., the spatial correlation is stronger). The performance of the APL and PQL methods are both good.

Tables~\ref{tab:Model selection results based on simulated samples, increased number of eff paras.} and \ref{tab:Model selection results based on simulated samples, increased number of non-eff paras.} show the results of varying the number of coefficients. The \textit{mis.coef} in Table~\ref{tab:Model selection results based on simulated samples, increased number of eff paras.} is larger compared to Table~\ref{tab:Model selection results based on simulated samples.}. In Table~\ref{tab:Model selection results based on simulated samples, increased number of eff paras.}, the value of fixed-effect parameters is changed. Some of the values are quite small (e.g., $-0.1$), and increases the difficulty of picking the correct model. The weak covariate effects sometimes can not be captured by the algorithms. From Table~\ref{tab:Model selection results based on simulated samples, increased number of non-eff paras.}, we notice that the variable selection performance is not affected when the number of noise variables increases.

In general, it is seen that the proposed variable selection methods perform reasonably well for independent or correlated covariates, different settings of fixed-effect and random-effect parameters. In terms of variable selection, the performance of the APL and the PQL are comparable. However, the PQL method requires less computing time when compared to the APL method. Supplementary Table~1 provides the computing time for one trial corresponding to the scenarios in Table~\ref{tab:Model selection results based on simulated samples, increased number of non-eff paras.}. While the computing time varies from case to case, we can see in general that the PQL method is about five times faster than the APL method.

\begin{table}
\centering
\caption{Model selection results based on simulated samples. The parameters are $\betavec=(-0.5, 0.75, 1, -0.75, -1, \zerovec_{10})'$, and $\thetavec=(0.1, 5)'$.}
\label{tab:Model selection results based on simulated samples.}
\vspace{1ex}
\begin{tabular}{r|r|rrr|rrr}
  \hline\hline
\multirow{2}{*}{Method} &\multirow{2}{*}{Cases}  & \multicolumn{3}{c|}{Consider spatial correlation} & \multicolumn{3}{c}{Ignore spatial correlation} \\ \cline{3-8}
 & &\textit{aver.size}  & \textit{corr.coef} & \textit{mis.coef} &\textit{aver.size}  & \textit{corr.coef} & \textit{mis.coef} \\\hline
 True value & & 5 & 10 & 0  & 5 & 10 & 0 \\\hline
 \multirow{5}{*}{APL.AEN}
 &Case 1 & 5.04 & 9.96 & 0.00 & 5.21 & 9.79 & 0.00 \\
 &Case 2 & 4.86 & 9.92 & 0.22 & 5.41 & 9.56 & 0.02 \\
 &Case 3 & 5.01 & 9.97 & 0.02 & 5.30 & 9.70 & 0.00 \\
 &Case 4 & 5.02 & 9.97 & 0.02 & 5.30 & 9.70 & 0.00 \\
 &Case 5 & 4.74 & 9.96 & 0.30 & 5.37 & 9.60 & 0.02 \\   \hline
\multirow{5}{*}{PQL.AEN}
 &Case 1 & 5.27 & 9.73 & 0.00 & 5.35 & 9.65 & 0.00 \\
 &Case 2 & 5.36 & 9.55 & 0.09 & 5.65 & 9.32 & 0.03 \\
 &Case 3 & 5.27 & 9.73 & 0.00 & 5.58 & 9.41 & 0.00 \\
 &Case 4 & 5.24 & 9.76 & 0.00 & 5.48 & 9.52 & 0.00 \\
 &Case 5 & 5.65 & 9.31 & 0.04 & 5.63 & 9.35 & 0.03 \\  \hline\hline
\end{tabular}
\end{table}

\begin{table}
\centering
\caption{Model selection results based on simulated samples. The parameters are $\betavec=(-0.5, 0.75, 1, -0.75, -1, \zerovec_{10})'$, and $\thetavec=(0.5, 5)'$.}
\label{tab:Model selection results based on simulated samples, increase sigma_sq.}
\vspace{1ex}
\begin{tabular}{r|r|rrr|rrr}
  \hline\hline
\multirow{2}{*}{Method} &\multirow{2}{*}{Cases}  & \multicolumn{3}{c|}{Consider spatial correlation} & \multicolumn{3}{c}{Ignore spatial correlation} \\ \cline{3-8}
 & &\textit{aver.size}  & \textit{corr.coef} & \textit{mis.coef} &\textit{aver.size}  & \textit{corr.coef} & \textit{mis.coef} \\\hline
 True value & & 5 & 10 & 0  & 5 & 10 & 0 \\\hline
 \multirow{5}{*}{APL.AEN}
   &Case 1 & 5.01 & 9.99 & 0.00 & 5.38 & 9.62 & 0.00 \\
   &Case 2 & 4.41 & 9.96 & 0.63 & 5.40 & 9.42 & 0.18 \\
   &Case 3 & 4.87 & 9.99 & 0.15 & 5.33 & 9.59 & 0.08 \\
   &Case 4 & 4.89 & 9.96 & 0.14 & 5.30 & 9.61 & 0.09 \\
   &Case 5 & 4.48 & 9.96 & 0.56 & 5.50 & 9.38 & 0.12 \\  \hline
\multirow{5}{*}{PQL.AEN}
   &Case 1 & 5.53 & 9.47 & 0.00 & 6.34 & 8.66 & 0.00 \\
   &Case 2 & 5.20 & 9.63 & 0.17 & 5.66 & 9.15 & 0.19 \\
   &Case 3 & 5.88 & 9.11 & 0.01 & 6.85 & 8.13 & 0.02 \\
   &Case 4 & 5.69 & 9.27 & 0.04 & 6.65 & 8.30 & 0.05 \\
   &Case 5 & 5.26 & 9.59 & 0.15 & 5.65 & 9.23 & 0.12 \\  \hline\hline
\end{tabular}
\end{table}

\begin{table}
\centering
\caption{Model selection results based on simulated samples. The parameters are $\betavec=(-0.5, 0.75, 1, -0.75, -1, \zerovec_{10})'$, and $\thetavec=(0.1, 10)'$.}
\label{tab:Model selection results based on simulated samples, increased d.}
\vspace{1ex}
\begin{tabular}{r|r|rrr|rrr}
  \hline\hline
\multirow{2}{*}{Method} &\multirow{2}{*}{Cases}  & \multicolumn{3}{c|}{Consider spatial correlation} & \multicolumn{3}{c}{Ignore spatial correlation} \\ \cline{3-8}
 & &\textit{aver.size}  & \textit{corr.coef} & \textit{mis.coef} &\textit{aver.size}  & \textit{corr.coef} & \textit{mis.coef} \\
  \hline
 True value & & 5 & 10 & 0  & 5 & 10 & 0 \\\hline
 \multirow{5}{*}{APL.AEN}
  &Case 1 & 5.02 & 9.98 & 0.00 & 5.25 & 9.75 & 0.00 \\
  &Case 2 & 4.86 & 9.95 & 0.19 & 5.37 & 9.62 & 0.00 \\
  &Case 3 & 5.03 & 9.96 & 0.02 & 5.26 & 9.74 & 0.00 \\
  &Case 4 & 5.03 & 9.94 & 0.02 & 5.26 & 9.74 & 0.00 \\
  &Case 5 & 4.74 & 9.97 & 0.29 & 5.29 & 9.67 & 0.03 \\  \hline
\multirow{5}{*}{PQL.AEN}
  &Case 1 & 5.20 & 9.80 & 0.00 & 5.35 & 9.65 & 0.00 \\
  &Case 2 & 5.43 & 9.50 & 0.06 & 5.61 & 9.38 & 0.01 \\
  &Case 3 & 5.25 & 9.74 & 0.01 & 5.47 & 9.53 & 0.00 \\
  &Case 4 & 5.26 & 9.74 & 0.00 & 5.45 & 9.55 & 0.00 \\
  &Case 5 & 5.31 & 9.60 & 0.09 & 5.36 & 9.60 & 0.04 \\   \hline\hline \end{tabular}
\end{table}

\begin{table}
\centering
\caption{Model selection results based on simulated samples. The parameters are $\betavec=(0.2, 0.3, 0.4, 0.5, 0.7, 0.8, -0.1, -0.6, -0.9, -1, \zerovec_{10})'$, and $\thetavec=(0.1, 5)'$.}
\label{tab:Model selection results based on simulated samples, increased number of eff paras.}
\vspace{1ex}
\begin{tabular}{r|r|rrr|rrr}
  \hline\hline
\multirow{2}{*}{Method} &\multirow{2}{*}{Cases}  & \multicolumn{3}{c|}{Consider spatial correlation} & \multicolumn{3}{c}{Ignore spatial correlation} \\ \cline{3-8}
 & &\textit{aver.size}  & \textit{corr.coef} & \textit{mis.coef} &\textit{aver.size}  & \textit{corr.coef} & \textit{mis.coef} \\
  \hline
True value & & 10 & 10 & 0  & 10 & 10 & 0 \\\hline
 \multirow{5}{*}{APL.AEN}
  &Case 1 & 9.08 & 9.97 & 0.95 & 10.10 & 9.43 & 0.48 \\
  &Case 2 & 8.97 & 9.97 & 1.06 & 9.60 & 9.65 & 0.75 \\
  &Case 3 & 8.92 & 9.93 & 1.15 & 9.64 & 9.57 & 0.79 \\
  &Case 4 & 8.72 & 9.97 & 1.31 & 9.56 & 9.60 & 0.84 \\
  &Case 5 & 8.84 & 9.95 & 1.20 & 9.27 & 9.68 & 1.05 \\ \hline
\multirow{5}{*}{PQL.AEN}
  &Case 1 & 9.76 & 9.68 & 0.55 & 10.19 & 9.33 & 0.47 \\
  &Case 2 & 9.79 & 9.64 & 0.58 & 10.08 & 9.27 & 0.65 \\
  &Case 3 & 9.62 & 9.69 & 0.69 & 9.99 & 9.32 & 0.69 \\
  &Case 4 & 9.55 & 9.65 & 0.80 & 10.00 & 9.26 & 0.74 \\
  &Case 5 & 9.47 & 9.64 & 0.89 & 9.95 & 9.13 & 0.92 \\   \hline
 \hline
\end{tabular}
\end{table}

\begin{table}
\centering
\caption{Model selection results based on simulated samples. The parameters are $\betavec=(-0.5, 0.75, 1, -0.75, -1, \zerovec_{20})'$, and $\thetavec=(0.1, 5)'$.}
\label{tab:Model selection results based on simulated samples, increased number of non-eff paras.}
\vspace{1ex}
\begin{tabular}{r|r|rrr|rrr}
  \hline\hline
\multirow{2}{*}{Method} &\multirow{2}{*}{Cases}  & \multicolumn{3}{c|}{Consider spatial correlation} & \multicolumn{3}{c}{Ignore spatial correlation} \\ \cline{3-8}
 & &\textit{aver.size}  & \textit{corr.coef} & \textit{mis.coef} &\textit{aver.size}  & \textit{corr.coef} & \textit{mis.coef} \\
  \hline
 True value & & 5 & 20 & 0  & 5 & 20 & 0 \\\hline
 \multirow{5}{*}{APL.AEN}
   &Case 1 & 5.05 & 19.95 & 0.00 & 5.29 & 19.71 & 0.00 \\
   &Case 2 & 4.82 & 19.85 & 0.32 & 5.60 & 19.35 & 0.06 \\
   &Case 3 & 5.04 & 19.94 & 0.02 & 5.41 & 19.59 & 0.00 \\
   &Case 4 & 5.04 & 19.93 & 0.03 & 5.59 & 19.41 & 0.00 \\
   &Case 5 & 4.83 & 19.88 & 0.29 & 5.56 & 19.39 & 0.04 \\\hline
  \multirow{5}{*}{PQL.AEN}
    &Case 1 & 5.24 & 19.76 & 0.00 & 5.54 & 19.46 & 0.00 \\
    &Case 2 & 5.17 & 19.70 & 0.14 & 5.65 & 19.28 & 0.07 \\
    &Case 3 & 5.40 & 19.59 & 0.00 & 5.86 & 19.13 & 0.01 \\
    &Case 4 & 5.47 & 19.53 & 0.00 & 6.06 & 18.94 & 0.00 \\
    &Case 5 & 5.28 & 19.62 & 0.10 & 5.51 & 19.43 & 0.06 \\ \hline\hline
\end{tabular}
\end{table}

%%%%%%%%%%%%%%%%%%%%%%%%%%%%%%%%%%%%%%%%%%%%%%%%%%%%%%%%%%%%%%%%%%%%%%%%%%%%%%%%%%%%%%%%%%%%%%%%%%%%%%%%%%%%%%%%%
\subsection{Comparisons with Existing Methods}
%%%%%%%%%%%%%%%%%%%%%%%%%%%%%%%%%%%%%%%%%%%%%%%%%%%%%%%%%%%%%%%%%%%%%%%%%%%%%%%%%%%%%%%%%%%%%%%%%%%%%%%%%%%%%%%%%
In this section, we extend the simulation studies in Section ~\ref{sec:simulation settings} to make comparisons with existing methods. Specifically, we compare the performance of the P-value-based method, the backward selection method, and the glmmLasso method as implemented in \citeN{glmmLasso}. Here, we briefly describe the three existing methods. For the P-value-based method, we use the R function \texttt{glmmPQL()} in \citeN{MASS} to fit the GLMM and obtain the p-value for each covariate. A covariate will stay in the model if its corresponding p-value is less than 0.05. For the backward selection method, we first use the \texttt{glmmPQL()} function to fit a full model. Then we do a backward elimination until all remaining covariates are significant (i.e., the p-value is less than 0.05). In each round, we eliminate the one with the highest p-value. For the glmmLasso method, we use the \texttt{glmmLasso()} function in the ``glmmLasso'' package (\citeNP{glmmLasso}). We first fit a GLMM model to generate the initial values then use the BIC to select the best penalty parameter.

For the three existing methods, we repeated all simulation settings as in Tables~\ref{tab:Model selection results based on simulated samples.}-\ref{tab:Model selection results based on simulated samples, increased number of non-eff paras.}. Here we discuss the comparison of the proposed and existing methods for the setting regarding to Table~\ref{tab:Model selection results based on simulated samples, increased number of non-eff paras.}. Table~\ref{tab:comp.existing} shows the model selection results for the three existing methods. The rest of the results are available in Supplementary Tables~2-5. Here, \textit{aver.size}, \textit{corr.coef}, and \textit{mis.coef} are abbreviated as ``AS'', ``CC'', and ``MC'', respectively. From Tables~\ref{tab:Model selection results based on simulated samples, increased number of non-eff paras.} and~\ref{tab:comp.existing}, the proposed APL and PQL work well with \textit{corr.coef} very close to 20 (the target is 20) and the \textit{mis.coef} is very close to zero. The P-value-based and backward methods work somewhat worse because the \textit{corr.coef} is around 18.5. For the glmmLasso method, the \textit{mis.coef} tends to be larger. We also observe a similar pattern for additional results in Supplementary Tables 2-5.  Overall, the proposed methods have advantages in variable selection under the setting of spatial variable selection with correlated covariates.

\begin{table}
\centering
\caption{Comparisons with existing methods for model selection results based on simulated samples. The setting is the same as in Table~\ref{tab:Model selection results based on simulated samples, increased number of non-eff paras.}. The metrics used are \textit{aver.size} (AS), \textit{corr.coef} (CC), and \textit{mis.coef} (MC).}\label{tab:comp.existing}
\vspace{.5em}
\begin{tabular}{r|rrr|rrr|rrr}\hline\hline
\multirow{2}{*}{Cases} &\multicolumn{3}{c|}{P-value-based}  & \multicolumn{3}{c|}{Backward} & \multicolumn{3}{c}{glmmLasso}\\\cline{2-10}
      & AS & CC & MC & AS & CC & MC & AS & CC & MC \\\hline
True value& 5   &   20 &    0 &    5 &   20 &     0&    5 &   20 &     0\\\hline
  Case 1 & 6.64 & 18.36 & 0.00 & 6.68 & 18.32 & 0.00 & 6.66 & 17.52 & 0.82 \\
  Case 2 & 6.21 & 18.78 & 0.01 & 6.41 & 18.58 & 0.01 & 3.16 & 19.75 & 2.09 \\
  Case 3 & 6.75 & 18.25 & 0.00 & 6.69 & 18.31 & 0.00 & 7.71 & 16.53 & 0.75 \\
  Case 4 & 6.87 & 18.13 & 0.00 & 6.82 & 18.18 & 0.00 & 7.57 & 16.69 & 0.74 \\
  Case 5 & 6.18 & 18.81 & 0.00 & 6.35 & 18.65 & 0.00 & 3.15 & 19.63 & 2.22 \\
  \hline\hline
\end{tabular}
\end{table}

%%%%%%%%%%%%%%%%%%%%%%%%%%%%%%%%%%%%%%%%%%%%%%%%%%%%%%%%%%%%%%%%%%%%%%%%%%%%%%%%%%%%%%%%%%%%%%%%%%%%%%%%%%%%%%%%%
\subsection{Comparison with Covariates Simulated from Real Data}
%%%%%%%%%%%%%%%%%%%%%%%%%%%%%%%%%%%%%%%%%%%%%%%%%%%%%%%%%%%%%%%%%%%%%%%%%%%%%%%%%%%%%%%%%%%%%%%%%%%%%%%%%%%%%%%%%
In this section, we consider a simulation scenario in which the covariates are sampled from the Virginia Lyme disease data. The details of the data analysis is given in Section~\ref{sec:application to Lyme disease data}. We use the parameter estimates of $\betavec$ and $\thetavec$ from Ecoregion 0 as the true values of the parameters in the simulation. For each simulated trial, we sample $n=225$ rows from the model matrix $X$ to obtain the covariate information. We then use the fitted model to simulate the number of counts. With the simulated data, we apply the proposed and existing methods to do variable selection. Similar to other settings, we repeat for 300 trials.

Table~\ref{tab:realX.sim} shows the comparisons of the proposed and existing methods for model selection results using covariates sampled from the real data. From the results, we can see that both the APL and PQL methods have the top two largest \textit{corr.coef}, while the backward selection and PQL methods are with the first and second smallest \textit{mis.coef}. We also notice that the \textit{mis.coef} is large for all methods. This is because there are three covariates with relatively small effect size, which is a challenging case for variable selection and thus the \textit{mis.coef} tends to be large. Overall, the PQL method has the best performance for this simulation scenario.

\begin{table}
\centering
\caption{Comparisons of the proposed and existing methods for model selection results using covariates sampled from the Lyme disease data.  The parameters are $\betavec=(0.503, 0.185, -0.161, 0.064, -0.048, 0.009, \zerovec_{8})'$, and $\thetavec=(0.417, 39.660)'$.}\label{tab:realX.sim}
\vspace{.5em}
\begin{tabular}{r|rrr}\hline\hline
 Methods & \textit{aver.size} & \textit{corr.coef} & \textit{mis.coef} \\ \hline
 True value      & 6    & 8    & 0    \\\hline
  APL.AEN        & 1.48 & 7.59 & 4.93 \\
  PQL.AEN        & 2.78 & 7.05 & 4.17 \\
  P-value-based  & 1.72 & 6.46 & 4.83 \\
  Backward       & 3.39 & 5.73 & 3.87 \\
  glmmLasso      & 1.04 & 6.41 & 5.54 \\\hline\hline
\end{tabular}
\end{table}

%%%%%%%%%%%%%%%%%%%%%%%%%%%%%%%%%%%%%%%%%%%%%%%%%%%%%%%%%%%%%%%%%%%%%%%%%%%%%%%%%%%%%%%%%%%%%%%%%%%%%%%%%%%%%%%%%
\section{Virginia Lyme Disease Data Analysis} \label{sec:application to Lyme disease data}
%%%%%%%%%%%%%%%%%%%%%%%%%%%%%%%%%%%%%%%%%%%%%%%%%%%%%%%%%%%%%%%%%%%%%%%%%%%%%%%%%%%%%%%%%%%%%%%%%%%%%%%%%%%%%%%%%
In this section, we present the data analysis for the Virginia Lyme disease data. To fit the GLMM to the Lyme disease data, we consider an exponential correlation function. We apply the PQL.AEN algorithm as described in Section~\ref{sec:estimation} to the Lyme disease data because of its computational efficiency.

We fit separate models to the two subregions in Virginia because the two subregions have different environmental and demographic characteristics, which could lead to different sets of active variables for the model and different spatial correlation patterns. The Subregion~0 ($n=583$), which consists of Northern Piedmont, Blue Ridge, Ridge and Valley and Central Appalachian areas, reported larger number of Lyme disease cases than the Subregion~1 ($n=1275$). Table~\ref{tab:Table of coefficients} lists the selected covariates, estimates of corresponding regression coefficients, and the estimates of parameters in the covariance structure. The results show that the factors that affect the Lyme disease case counts are different for the two subregions. Here, we interpret the selected variables for each subregion.

For Subregion~0 (i.e., the northern/western sub-region), the selected variables are percentage of forest (Forest\_NLCD06), percentage of herbaceous (Herbaceous\_NLCD06), developed-forest edge (TECI\_DF06), forest-herbaceous edge (TECI\_FH06), population density (Pop\_den), and mean income (Mean\_income). In particular,

\begin{inparaitem}
\item  the percentage of herbaceous cover has positive relationships with Lyme disease case counts, which is consistent with the findings in \citeN{Jackson2006}. Herbaceous (especially scrub) areas can provide favorable living environment for deer and mice.

\item The forest-herbaceous edge is positively correlated with the disease counts. The mixture of forest and herbaceous areas is appealing for some host animals. For example, deer always stays within a short distance of forest cover (forest edge), but forest cover provides too much shade to grow many of the plants that deer like to feed on whereas scrub offers more sunlight for vegetative growth while still providing some cover. Therefore, the interspersion of forest and herbaceous land can have a positive relationship with Lyme disease incidence.

\item Although the percentage of forest cover is negatively correlated with the case counts, we notice that the correlation between Forest\_NLCD06 and TECI\_FH06 is 0.89. Because the forest-herbaceous edge has a strong positive effect, the combined effects of the forest cover and forest-herbaceous edge can still be positive. The percentage of forest cover was also found out to be an important variable in literature (\citeNP{Jackson2006}).

\item The developed-forest edge is positively correlated with the case counts. Due to development, forest communities were fragmented by suburban, creating developed-forest edge. The developed-forest edge results in a habitat environment that is suitable for deer, small rodents, and the white-footed mouse (e.g., Page 150 of \citeNP{MayerPizer2008}). Those animals contribute to the hosting and transmission of the Lyme disease.

\item The population density is negatively correlated with the case counts. This is because large population density often means that the area is an  urban and developed regions. In those regions, the environments tend to provide fewer habitats for ticks and disease reservoirs. There is also less human-environment interaction in highly populated regions.

\item The mean income was also found out to be an active variable, which is consistent with the Lyme disease literature. Both \citeN{Jackson2006} and \shortciteN{Seukepetal2015} found that income is correlated with Lyme disease incidence. As pointed out by \shortciteN{Seukepetal2015}, counties in northern Virginia (outside Washington DC) are wealthy, and Lyme disease incidence rates have been high in that region, which likely contributes to the correlation between income and incidence rates.

\end{inparaitem}

For Subregion~1 (i.e., the southern/eastern sub-region), the selected variables are percentage of developed land (Dvlpd\_NLCD06), forested fragments (Tract\_Frag06), developed-forest edge (CWED\_DF06), herbaceous-developed edge (TECI\_HD06), median age (Median\_age), and mean income (Mean\_income). In addition to those variables already interpreted in Ecoregion~0,

\begin{inparaitem}
\item the percent developed has a negative correlation with the number of Lyme disease cases. For areas with high percentage of development such as in inner cities, the population is not exposed to much Lyme because that environment does not support deer or white-footed mouse habitat. Areas with lower percentage than inner cities, such as many suburban counties, tend to have a lot of deer in and around residential areas. In Subregion~1, Lyme disease incidence appears to be more prevalent in counties that have some suburban sprawl (e.g., the counties around Richmond City or Lynchburg City, or the counties between Richmond and the cities of the Hampton Roads along Interstate Highway-64).

\item The forested fragments is negatively correlated with the case counts. Because the correlation between Tract\_Frag06 and CWED\_DF06 is 0.60, and the developed-forest edge has a strong positive effect, the combined effects of the forested fragments and developed-forest edge can still be positive. The forested fragments was also found out to be an important variable in literature (\citeNP{Jackson2006}).

\item The presence of interspersion of herbaceous and developed areas has a negative correlation with Lyme disease incidence. The present of interspersion of herbaceous and developed limit the movement of white-footed mice into the developed area.  White footed mice are the primary contributors to the dispersion of infected larval-stage ticks, which then develop into nymph stage ticks that bite people and transmit Lyme.  However, white-footed mice are primarily a forest species and while they may spend time in scrub habitats near a forest tract they might not venture far from the forest edge.

\item Median age is positively correlated with the case counts. Census tracts with older populations tend to have higher incidence rates because Lyme disease tends to appear more in adults older than 40 (e.g., \citeNP{KilpatrickLaBonte2007}).

\end{inparaitem}

Table~\ref{tab:Table of coefficients} also shows the corresponding approximate 95\% bootstrap CIs for parameters based on $B=1000$ bootstrap samples. For Ecoregion~0, the variable with CI excludes zero is TECI\_FH06, and for Ecoregion~1, the variables with CIs exclude zero are CWED\_DF06, Median\_age, and Mean\_income. The results indicate that environmental variables that are related to edges (i.e., developed-forest edge and forest-herbaceous edge) and the income variable are particularly important for the disease emergency. We also note that, for Ecoregion~0, the CIs for the regression coefficients of Dvlpd\_NLCD06 and Forest\_NLCD06 are wide, due to the strong correlation between the two variables (i.e., the correlation is $-$0.85).

The results in this paper are largely consistent with results in \citeN{Allan2003}, \citeN{Jackson2006}, and \shortciteN{Seukepetal2015} but with new findings. We found that the developed-forest edge and forest-herbaceous edge are particularly important for the Lyme disease counts. In our study, the forest fragment perimeters (FragPerim06) is not included in the final models for either ecoregions. \shortciteN{Seukepetal2015} found that the percentage of forest was not selected, while our findings support the results in \citeN{Jackson2006} and suggest that it is an important variable.

\shortciteN{Seukepetal2015} fitted a spatial model using Lyme disease data without considering the ecoregion variable. We show that two ecoregions have different sets of active variables. In addition, the estimates of parameters in the covariance structure are different in the two ecoregions. Specifically, the estimated scale parameter $d$ in the correlation function is quite small in Subregion~1, which implies that the spatial correlation is weak in that subregion. For subregion~0, the estimated $d$ is 39.660. That is, when the distance between two census tracts is 39.66 kilometer (km), the correlation is estimated to be 0.37.  The estimated $\sigmahat^2$ in two subregions are close.

\begin{table}
\centering
\caption{The list of selected covariates, estimates of corresponding regression coefficients, the estimates of parameters in the covariance structure, and their corresponding approximate 95\% bootstrap CIs. Note that separate models were fitted for the two subregions.} \label{tab:Table of coefficients}
\vspace{1ex}
\begin{tabular}{c|rrrc|rrr}\hline\hline
\multirow{3}{*}{Parameters}         & \multicolumn{3}{c}{Ecoregion 0} & &  \multicolumn{3}{c}{Ecoregion 1} \\\cline{2-8}
    & \multirow{2}{*}{estimate} & \multicolumn{2}{c}{95\% CI}&& \multirow{2}{*}{estimate}  & \multicolumn{2}{c}{95\% CI} \\\cline{3-4}\cline{7-8}
    & & lower & upper && & lower & upper\\\hline
  Intercept          &    1.794 &   1.012 & 2.543 &&    1.393 &   1.297 & 1.480 \\
  Dvlpd\_NLCD06      &        0 &       0 & 5.095 && $-$0.115 &$-$0.469 &     0 \\
  Forest\_NLCD06     & $-$0.161 &$-$3.999 &     0 &&        0 &       0 & 0.278 \\
  Herbaceous\_NLCD06 &    0.009 &$-$0.430 & 1.643 &&        0 &       0 & 0.174 \\
  Tract\_Frag06      &        0 &       0 & 1.362 && $-$0.057 &$-$0.416 &     0 \\
  FragPerim06        &        0 &       0 & 1.600 &&        0 &       0 & 0.416 \\
  CWED\_DF06         &        0 &       0 & 0.377 &&    0.173 &   0.096 & 0.507 \\
  TECI\_DF06         &    0.064 &       0 & 0.520 &&        0 &       0 & 0.292 \\
  CWED\_FH06         &        0 &       0 & 0.481 &&        0 &       0 & 0.181 \\
  TECI\_FH06         &    0.503 &   0.297 & 1.047 &&        0 &       0 & 0.395 \\
  CWED\_HD06         &        0 &       0 & 0.383 &&        0 &       0 & 0.380 \\
  TECI\_HD06         &        0 &       0 & 0.419 && $-$0.231 &$-$0.421 &     0 \\
  Pop\_den           & $-$0.048 &$-$0.914 &     0 &&        0 &       0 & 0.124 \\
  Median\_age        &        0 &       0 & 0.449 &&    0.136 &   0.028 & 0.248 \\
  Mean\_income       &    0.185 &       0 & 0.372 &&    0.397 &   0.347 & 0.487 \\\hline
  $\sigma^2$         &    0.417 &   0.248 & 0.889 &&    0.457 &   0.392 & 0.524 \\
$d$ (in km)          &   39.660 &   0.026 &64.402 &&    1.314 &   0.043 & 6.397 \\\hline\hline
\end{tabular}
\end{table}

%%%%%%%%%%%%%%%%%%%%%%%%%%%%%%%%%%%%%%%%%%%%%%%%%%%%%%%%%%%%%%%%%%%%%%%%%%%%%%%%%%%%%%%%%%%%%%%%%%%%%%%%%%%%%%%%%
\section{Conclusions and Areas for Future Research} \label{sec:conclusion}
%%%%%%%%%%%%%%%%%%%%%%%%%%%%%%%%%%%%%%%%%%%%%%%%%%%%%%%%%%%%%%%%%%%%%%%%%%%%%%%%%%%%%%%%%%%%%%%%%%%%%%%%%%%%%%%%%
In this paper, we consider the problem of variable selection in the spatial Poisson regression. By using the AEN penalty, we perform variable selection and parameter estimation simultaneously. We consider both APL and PQL methods for parameter estimations. Simulation studies in Section \ref{sec:simulation study} show that both methods perform reasonably well and their performance are comparable to each other. The comparisons with existing methods show that the developed methods have advantages in the setting of spatial variable selection. We then apply our method to select important variables associated with the Lyme disease emergence in Virginia.

For the Lyme disease research community, we develop an automatic variable selection procedure while accounting for spatial correlation. We used statewide Lyme disease data and covariates at census tract level to identify important environmental and human factors, which is new to the literature. Interestingly, we found different ecoregions have different sets of factors that are important to the disease spread, which can be important for disease monitoring.

In our analysis, we use datasets from different resources with different collection frequencies. For example, the US census data are updated every ten years, and the land cover data are updated every five years. Because our explanatory variables are aggregated at census tract level, we expect that the temporal changes over a five-year span to be a second order. For another perspective, the life cycle of ticks that causes Lyme disease is two years. Using a study period of five years allows us to study the overall effects of environmental and economic variables on the Lyme disease occurrence. However, we do want to point out that the temporal misalignment in Lyme disease counts and covariates could be one limitation of this study.

In disease mapping applications, it is not uncommon to have ``nugget'' effects (i.e., an unstructured Gaussian random effect). For our Lyme disease application, we did some model checking to see if it is necessary to add an unstructured Gaussian random effect. We computed the estimated number of counts for each census tract and plot it versus the observed number of counts. The results are shown in Supplementary Figure~2. From the plot, we can see most points align well with the 45-degree line. The overall $\textrm{R}^2$ is 96.1\%, indicating that the model can explain most of the variation in the data. Thus it is not necessary to add an unstructured term in our model. However, spatial variable selection with nugget effects could be an interesting topic for future research.

In this paper, we use Laplace approximation to the integrals in likelihood functions. As for future research, Bayesian methods can also be used as alternative to approximate integrals. We may use Gibbs sampler, Metropolis-Hastings algorithm, Markov chain Monte Carlo, importance sampling, to name a few. However, this is usually time-consuming. Also, we consider a Poisson regression model with random effects and a dispersion parameter $\phi$ equal to one. If over-dispersion appears in the data, we can add the dispersion parameter into model formulation and obtain estimates of $\{\betavec, \thetavec, \phi\}$ simultaneously. In some cases, one may encounter a dataset with large $n$. \citeN{KaufmanSchervishNychka2008} developed the covariance tapering method for large irregularly spaced data or missing data on lattice. By taking the inner product of a covariance matrix with a positive definite and compactly supported correlation matrix, one can obtain the ``tapered'' covariance matrix with sparsity. Future research can be devoted to incorporating the covariance tapering method for large $n$ case to achieve computational efficiency. In this paper, we use parametric bootstrap as a practical way to construct CIs. It would be an interesting topic for future research to develop a theoretical framework for CIs under the spatial variable selection setting.

%\iffalse
%%%%%%%%%%%%%%%%%%%%%%%%%%%%%%%%%%%%%%%%%%%%%%%%%%%%%%%%%%%%%%%%%%%%%%%%%%%%%%%%%%%%%%%%%%%%%%%%%%%%%%%%%%%%%%%%%
\section*{Acknowledgments}\label{Acknowledgement}
%%%%%%%%%%%%%%%%%%%%%%%%%%%%%%%%%%%%%%%%%%%%%%%%%%%%%%%%%%%%%%%%%%%%%%%%%%%%%%%%%%%%%%%%%%%%%%%%%%%%%%%%%%%%%%%%%
The authors would like to thank the editor, an associate editor, two referees, and an associate editor for reproducibility, for their valuable comments that helped in improving this paper significantly. The authors acknowledge Advanced Research Computing at Virginia Tech for providing computational resources. The research by Xie, Li, Kolivras, and Gaines was supported by National Science Foundation Grant BCS-1122876 to Virginia Tech. The research by Hong and Xu was partially supported by National Science Foundation Grants BCS-1122876 and CNS-1565314 to Virginia Tech.

%\fi

%%%%%%%%%%%%%%%%%%%%%%%%%%%%%%%%%%%%%%%%%%%%%%%%%%%%%%%%%%%%%%%%%%%%%%%%%%%%%%%%%%%%%%%%%%%%%%%%%%%%%%%%%%%%%%%%%
\appendix
%%%%%%%%%%%%%%%%%%%%%%%%%%%%%%%%%%%%%%%%%%%%%%%%%%%%%%%%%%%%%%%%%%%%%%%%%%%%%%%%%%%%%%%%%%%%%%%%%%%%%%%%%%%%%%%%%
\section{Quadratic Approximation to PQL} \label{sec:quadratic approximation}
%%%%%%%%%%%%%%%%%%%%%%%%%%%%%%%%%%%%%%%%%%%%%%%%%%%%%%%%%%%%%%%%%%%%%%%%%%%%%%%%%%%%%%%%%%%%%%%%%%%%%%%%%%%%%%%%%
Given current estimates of $\thetavec$ and $\bvec$, which are denoted by $\thetavectilde$ and $\bvectilde$, respectively, \eqref{eqn:la.equation} reduces to
\begin{align} \label{eqn:objection PQL function of beta }
l_a(\betavec|\thetavectilde, \bvectilde) = \sum\limits_{i=1}^{n}\left(-\mu_i+y_i \xvec_i'\betavec \right),
\end{align}
up to a constant that is independent of $\betavec$. We apply a quadratic approximation to $l_{a}(\betavec|\thetavectilde, \bvectilde)$ around the current estimate $\betavectilde$. That is
$$
l_{a}(\betavec|\thetavectilde, \bvectilde) \approx l_{a}(\betavectilde|\thetavectilde, \bvectilde)+ \dfrac{\partial l_{a}(\betavec|\thetavectilde, \bvectilde)}{\partial \betavec'}\bigg|_{\betavec=\betavectilde}(\betavec-\betavectilde)+\frac{1}{2} (\betavec-\betavectilde)' \dfrac{\partial^2 l_{a}(\betavec|\thetavectilde, \bvectilde)}{\partial \betavec \partial \betavec'} \bigg|_{\betavec=\betavectilde}(\betavec-\betavectilde),
$$
where
\begin{align*}
\dfrac{\partial l_{a}(\betavec|\thetavectilde, \bvectilde)}{\partial \betavec}\bigg|_{\betavec=\betavectilde} = \sum\limits_{i=1}^{n}\left(-\mu_i\xvec_i+y_i \xvec_i \right),\quad\textrm{ and }\quad
\dfrac{\partial^2 l_{a}(\betavec|\thetavectilde, \bvectilde)}{\partial \betavec \partial \betavec'}\bigg|_{\betavec=\betavectilde} = \sum\limits_{i=1}^{n}-\mu_i\xvec_i\xvec_i'
\end{align*}
are the first and second derivatives of $l_{a}(\betavec|\thetavectilde, \bvectilde)$ with respect to $\betavec$, respectively. Therefore,
\begin{align*}
l_{a}(\betavec|\thetavectilde, \bvectilde) & \approx \dfrac{\partial l_{a}(\betavec|\thetavectilde, \bvectilde)}{\partial \betavec'}\bigg|_{\betavec=\betavectilde}  \betavec +\frac{1}{2} (\betavec-\betavectilde)' \dfrac{\partial^2 l_{a}(\betavec|\thetavectilde, \bvectilde)}{\partial \betavec \partial \betavec'}\bigg|_{\betavec=\betavectilde} (\betavec-\betavectilde) + c\\
&= \sum\limits_{i=1}^{n} (y_i-\mu_i)\xvec_i'\betavec+\frac{1}{2} (\betavec-\betavectilde)' \left[ \sum\limits_{i=1}^{n} (-\mu_i \xvec_i \xvec_i') \right] (\betavec-\betavectilde)+ c \\
&=-\frac{1}{2}\sum\limits_{i=1}^{n}\mu_i\left[ 2\left(1-\frac{y_i}{\mu_i}\right) \xvec_i'\betavec + (\betavec-\betavectilde)'\xvec_i \xvec_i' (\betavec-\betavectilde)  \right] +c \\
&=-\frac{1}{2}\sum\limits_{i=1}^{n}\mu_i\left(\xvec_i'\betavectilde-1+\frac{y_i}{\mu_i}-\xvec_i'\betavec \right)^2 +c \\
&=-\frac{1}{2}\sum\limits_{i=1}^{n}\mu_i(z_i-\xvec_i'\betavec)^2 +c,
\end{align*}
where $z_i=\xvec_i'\betavectilde-1+y_{i}/\mu_{i}$ (i.e., the working response), $c$ is a constant that does not depend on $\betavec$.

%%%%%%%%%%%%%%%%%%%%%%%%%%%%%%%%%%%%%%%%%%%%%%%%%%%%%%%%%%%%%%%%%%%%%%%%%%%%%%%%%%%%%%%%%%%%%%%%%%%%%%%%%%%%%%%%%
\section*{Supplementary Materials}
%%%%%%%%%%%%%%%%%%%%%%%%%%%%%%%%%%%%%%%%%%%%%%%%%%%%%%%%%%%%%%%%%%%%%%%%%%%%%%%%%%%%%%%%%%%%%%%%%%%%%%%%%%%%%%%%%
The following supplementary materials are available online.
\begin{description}
\item[Additional details]  Additional computing and simulation results (pdf file).

\item[Data and code] The Virginia Lyme disease data and R code for simulation and analysis (zip file).

\item[R packages] The Virginia Lyme disease data and R code for algorithm implementation are also available in R packages ``VALymeData'' and ``SpatialVS'', respectively, which can be downloaded from the Comprehensive R Archive Network (CRAN), https://cran.r-project.org/. (R package).
\end{description}

%%%%%%%%%%%%%%%%%%%%%%%%% ref %%%%%%%%%%%%%%%%%%%%%%%%%%%%%%%%%%%%%%%%%%%%%%%%%%%%%%%%%%%%%%%%%%%%%%%%%%%%%%%%%%%%%%%%%%%%%%%%%
%\bibliographystyle{chicago}
%\bibliography{ref}

\begin{thebibliography}{}

\bibitem[\protect\citeauthoryear{Allan, Keesing, and Ostfeld}{Allan
  et~al.}{2003}]{Allan2003}
Allan, B.~F., F.~Keesing, and R.~S. Ostfeld (2003).
\newblock Effect of forest fragmentation on {Lyme} disease risk.
\newblock {\em Conservation Biology\/}~{\em 17}, 267--272.

\bibitem[\protect\citeauthoryear{Almquist}{Almquist}{2010}]{Almquist2010}
Almquist, Z.~W. (2010).
\newblock {US} census spatial and demographic data in {R}: The {UScensus2000}
  suite of packages.
\newblock {\em Journal of Statistical Software\/}~{\em 37}, 1--31.

\bibitem[\protect\citeauthoryear{Bachoc, Preinerstorfer, and
  Steinberger}{Bachoc et~al.}{2016}]{Bachocetal2016}
Bachoc, F., D.~Preinerstorfer, and L.~Steinberger (2016).
\newblock Uniformly valid confidence intervals post-model-selection.
\newblock {\em arXiv:1611.01043\/}.

\bibitem[\protect\citeauthoryear{Berk, Brown, Buja, Zhang, and Zhao}{Berk
  et~al.}{2013}]{berk2013valid}
Berk, R., L.~Brown, A.~Buja, K.~Zhang, and L.~Zhao (2013).
\newblock Valid post-selection inference.
\newblock {\em The Annals of Statistics\/}~{\em 41}, 802--837.

\bibitem[\protect\citeauthoryear{{Boehm Vock}, Reich, Fuentes, and
  Dominici}{{Boehm Vock} et~al.}{2015}]{BoehmVocketal2015}
{Boehm Vock}, L.~F., B.~J. Reich, M.~Fuentes, and F.~Dominici (2015).
\newblock Spatial variable selection methods for investigating acute health
  effects of fine particulate matter components.
\newblock {\em Biometrics\/}~{\em 71}, 167--177.

\bibitem[\protect\citeauthoryear{Breslow and Clayton}{Breslow and
  Clayton}{1993}]{BreslowClayton1993}
Breslow, N.~E. and D.~G. Clayton (1993).
\newblock Approximate inference in generalized linear mixed models.
\newblock {\em Journal of the American Statistical Association\/}~{\em 88},
  9--25.

\bibitem[\protect\citeauthoryear{Cai and Dunson}{Cai and
  Dunson}{2006}]{CaiDunson2006}
Cai, B. and D.~B. Dunson (2006).
\newblock Bayesian covariance selection in generalized linear mixed models.
\newblock {\em Biometrics\/}~{\em 62}, 446--457.

\bibitem[\protect\citeauthoryear{Chatterjee and Lahiri}{Chatterjee and
  Lahiri}{2011}]{chatterjee2011bootstrapping}
Chatterjee, A. and S.~N. Lahiri (2011).
\newblock Bootstrapping {Lasso} estimators.
\newblock {\em Journal of the American Statistical Association\/}~{\em 106},
  608--625.

\bibitem[\protect\citeauthoryear{Cui}{Cui}{2011}]{Cui2011}
Cui, R. (2011).
\newblock {\em Variable selection methods for longitudinal data}.
\newblock {PhD thesis}, Harvard University, Cambridge, MA.

\bibitem[\protect\citeauthoryear{Diggle, Moyeed, and Tawn}{Diggle
  et~al.}{1998}]{DiggleMoyeedTawn1998}
Diggle, P., R.~A. Moyeed, and J.~A. Tawn (1998).
\newblock Model-based geostatistics (with discussion).
\newblock {\em Journal of the Royal Statistical Society, Series C\/}~{\em 47},
  299--350.

\bibitem[\protect\citeauthoryear{{Ecoregion of Virginia}}{{Ecoregion of
  Virginia}}{2015}]{EcoregionVA}
{Ecoregion of Virginia} (2015).
\newblock Level {III} ecoregion map of {Virginia}.
\newblock
  \url{https://www.hort.purdue.edu/newcrop/cropmap/virginia/maps/VAeco3.html}.
\newblock Accessed: 2015-09-30.

\bibitem[\protect\citeauthoryear{Efron, Hastie, Johnstone, and
  Tibshirani}{Efron et~al.}{2004}]{Efronetal2004}
Efron, B., T.~Hastie, I.~Johnstone, and R.~Tibshirani (2004).
\newblock Least angle regression.
\newblock {\em The Annals of Statistics\/}~{\em 32}, 407--499.

\bibitem[\protect\citeauthoryear{Fahrmeir and Tutz}{Fahrmeir and
  Tutz}{2001}]{FahrmeirTutz2001}
Fahrmeir, L. and G.~Tutz (2001).
\newblock {\em Multivariate Statistical Modelling Based on Generalized Linear
  Models\/} (Second ed.).
\newblock New York: Springer.

\bibitem[\protect\citeauthoryear{Fan and Li}{Fan and Li}{2001}]{FanLi2001}
Fan, J. and R.~Li (2001).
\newblock Variable selection via nonconcave penalized likelihood and its oracle
  properties.
\newblock {\em Journal of the American Statistical Association\/}~{\em 96},
  1348--1360.

\bibitem[\protect\citeauthoryear{Fan and Lv}{Fan and Lv}{2010}]{FanLv2010}
Fan, J. and J.~Lv (2010).
\newblock A selective overview of variable selection in high dimensional
  feature space.
\newblock {\em Statistica Sinica\/}~{\em 20}, 101--148.

\bibitem[\protect\citeauthoryear{Friedman, Hastie, and Tibshirani}{Friedman
  et~al.}{2010}]{FriedmanHastieTibshirani2010}
Friedman, J., T.~Hastie, and R.~Tibshirani (2010).
\newblock Regularization paths for generalized linear models via coordinate
  descent.
\newblock {\em Journal of Statistical Software\/}~{\em 33}.

\bibitem[\protect\citeauthoryear{Fry, Xian, Jin, Dewitz, Homer, Yang, Barnes,
  Herold, and Wickham}{Fry et~al.}{2012}]{Fryetal2012}
Fry, J.~A., G.~Xian, S.~Jin, J.~A. Dewitz, C.~G. Homer, L.~Yang, C.~A. Barnes,
  N.~D. Herold, and J.~D. Wickham (2012).
\newblock Completion of the 2006 national land cover database update for the
  conterminous {United} {States}.
\newblock {\em Photogrammetric Engineering and Remote Sensing\/}~{\em 77},
  858--864.

\bibitem[\protect\citeauthoryear{Groll}{Groll}{2016}]{glmmLasso}
Groll, A. (2016).
\newblock {\em {glmmLasso}: Variable Selection for Generalized Linear Mixed
  Models by {$L_1$}-Penalized Estimation}.
\newblock R package version 1.4.4.

\bibitem[\protect\citeauthoryear{Groll and Tutz}{Groll and
  Tutz}{2014}]{GrollTutz2014}
Groll, A. and G.~Tutz (2014).
\newblock Variable selection for generalized linear mixed models by
  ${L}_1$-penalized estimation.
\newblock {\em Statistics and Computing\/}~{\em 24}, 137--154.

\bibitem[\protect\citeauthoryear{Hoerl and Kennard}{Hoerl and
  Kennard}{1970}]{HorelKennard1970}
Hoerl, A.~E. and R.~W. Kennard (1970).
\newblock Ridge regression: Biased estimation for nonorthogonal problems.
\newblock {\em Technometrics\/}~{\em 12}, 55--67.

\bibitem[\protect\citeauthoryear{Hong, Xu, Xie, and Jin}{Hong
  et~al.}{2018a}]{SpatialVS-Rpackage}
Hong, Y., L.~Xu, Y.~Xie, and Z.~Jin (2018a).
\newblock {\em SpatialVS: Spatial Variable Selection}.
\newblock R package version 1.0.

\bibitem[\protect\citeauthoryear{Hong, Xu, Xie, and Jin}{Hong
  et~al.}{2018b}]{SpatialVS-Datapackage}
Hong, Y., L.~Xu, Y.~Xie, and Z.~Jin (2018b).
\newblock {\em VALymeData: The Virginia Lyme Disease Data}.
\newblock R package version 1.0.

\bibitem[\protect\citeauthoryear{Ibrahim, Zhu, Garcia, and Guo}{Ibrahim
  et~al.}{2011}]{Ibrahimetal2011}
Ibrahim, J.~G., H.~Zhu, R.~I. Garcia, and R.~Guo (2011).
\newblock Fixed and random effects selection in mixed effects models.
\newblock {\em Biometrics\/}~{\em 67}, 495--503.

\bibitem[\protect\citeauthoryear{Jackson, Hilborn, and Thomas}{Jackson
  et~al.}{2006}]{Jackson2006}
Jackson, L.~E., E.~D. Hilborn, and J.~C. Thomas (2006).
\newblock Towards landscape design guidelines for reducing {Lyme} disease risk.
\newblock {\em International Journal of Epidemiology\/}~{\em 35}, 315--322.

\bibitem[\protect\citeauthoryear{Kaufman, Schervish, and Nychka}{Kaufman
  et~al.}{2008}]{KaufmanSchervishNychka2008}
Kaufman, C., M.~Schervish, and D.~Nychka (2008).
\newblock Covariance tapering for likelihood based estimation in large spatial
  datasets.
\newblock {\em Journal of the American Statistical Association\/}~{\em 103},
  1545--1555.

\bibitem[\protect\citeauthoryear{Kilpatrick and {LaBonte}}{Kilpatrick and
  {LaBonte}}{2007}]{KilpatrickLaBonte2007}
Kilpatrick, H.~J. and A.~M. {LaBonte} (2007).
\newblock Managing urban deer in {Connecticut}: a guide for residents and
  communities concerned about overabundant deer populations.
\newblock
  \url{http://www.ct.gov/deep/lib/deep/wildlife/pdf_files/game/urbandeer07.pdf}.
\newblock Accessed: 2018-05-20.

\bibitem[\protect\citeauthoryear{Laplace}{Laplace}{1986}]{Laplace1986}
Laplace, P.~S. (1986).
\newblock Memoir on the probability of the causes of events.
\newblock {\em Statistical Science\/}~{\em 1}, 364--378.

\bibitem[\protect\citeauthoryear{Lee, Sun, Sun, and Taylor}{Lee
  et~al.}{2016}]{lee2016exact}
Lee, J.~D., D.~L. Sun, Y.~Sun, and J.~E. Taylor (2016).
\newblock Exact post-selection inference, with application to the lasso.
\newblock {\em The Annals of Statistics\/}~{\em 44}, 907--927.

\bibitem[\protect\citeauthoryear{Li, Hong, Thapa, and Burkhart}{Li
  et~al.}{2015}]{Lietal2015}
Li, J., Y.~Hong, R.~Thapa, and H.~E. Burkhart (2015).
\newblock Survival analysis of loblolly pine trees with spatially correlated
  random effects.
\newblock {\em Journal of the American Statistical Association\/}~{\em 101},
  486--502.

\bibitem[\protect\citeauthoryear{Li, Kolivras, Hong, Duan, Seukep, Prisley,
  Campbell, and Gaines}{Li et~al.}{2014}]{Lietal2014}
Li, J., K.~N. Kolivras, Y.~Hong, Y.~Duan, S.~E. Seukep, S.~P. Prisley, J.~B.
  Campbell, and D.~N. Gaines (2014).
\newblock Spatial and temporal emergence pattern of {Lyme} disease in
  {Virginia}.
\newblock {\em The American Journal of Tropical Medicine and Hygiene\/}~{\em
  91}, 1166--1172.

\bibitem[\protect\citeauthoryear{Lu, Liu, Yin, and Zhang}{Lu
  et~al.}{2017}]{lu2017confidence}
Lu, S., Y.~Liu, L.~Yin, and K.~Zhang (2017).
\newblock Confidence intervals and regions for the lasso by using stochastic
  variational inequality techniques in optimization.
\newblock {\em Journal of the Royal Statistical Society: Series B\/}~{\em 79},
  589--611.

\bibitem[\protect\citeauthoryear{Maes, Lecomte, and Ray}{Maes
  et~al.}{1998}]{Maes1998}
Maes, E., P.~Lecomte, and N.~Ray (1998).
\newblock A cost-of-illness study of {L}yme disease in the {U}nited {S}tates.
\newblock {\em Clinical Therapeutics\/}~{\em 20}, 993--1008.

\bibitem[\protect\citeauthoryear{Mayer and Pizer}{Mayer and
  Pizer}{2008}]{MayerPizer2008}
Mayer, K.~H. and H.~F. Pizer (2008).
\newblock {\em The Social Ecology of Infectious Diseases}.
\newblock Burlington, MA: Academic Press.

\bibitem[\protect\citeauthoryear{McCulloch, Searle, and Neuhaus}{McCulloch
  et~al.}{2008}]{McCullochSearleNeuhaus2008}
McCulloch, C.~E., S.~R. Searle, and J.~M. Neuhaus (2008).
\newblock {\em Generalized, Linear and Mixed Models. 2nd Edition}.
\newblock New Jersey: John Wiley and Sons.

\bibitem[\protect\citeauthoryear{McGarigal, Cushman, and Ene}{McGarigal
  et~al.}{2012}]{McGarigaletal2012}
McGarigal, K., S.~A. Cushman, and E.~Ene (2012).
\newblock {\em {FRAGSTATS} v4: Spatial Pattern Analysis Program for Categorical
  and Continuous Maps}.
\newblock http://www.umass.edu/landeco/research/fragstats/fragstats.html:
  University of Massachusetts, Amherst.

\bibitem[\protect\citeauthoryear{Ning and Liu}{Ning and
  Liu}{2017}]{ning2017general}
Ning, Y. and H.~Liu (2017).
\newblock A general theory of hypothesis tests and confidence regions for
  sparse high dimensional models.
\newblock {\em The Annals of Statistics\/}~{\em 45}, 158--195.

\bibitem[\protect\citeauthoryear{O{'}Hara and Sillanp{\"a}{\"a}}{O{'}Hara and
  Sillanp{\"a}{\"a}}{2009}]{OharaSillanpaa2009}
O{'}Hara, R.~B. and M.~J. Sillanp{\"a}{\"a} (2009).
\newblock A review of {Bayesian} variable selection methods: What, how and
  which.
\newblock {\em Bayesian Analysis\/}~{\em 4}, 85--118.

\bibitem[\protect\citeauthoryear{Park and Hastie}{Park and
  Hastie}{2007}]{ParkHastie2007}
Park, M.~Y. and T.~Hastie (2007).
\newblock {$L_1$}-regularization path algorithm for generalized linear models.
\newblock {\em Journal of the Royal Statistical Society, Series B\/}~{\em 69},
  659--677.

\bibitem[\protect\citeauthoryear{{R Core Team}}{{R Core Team}}{2016}]{R2016}
{R Core Team} (2016).
\newblock {\em R: A Language and Environment for Statistical Computing}.
\newblock Vienna, Austria: R Foundation for Statistical Computing.

\bibitem[\protect\citeauthoryear{Schelldorfer, Meier, and
  B{\"u}hlmann}{Schelldorfer et~al.}{2012}]{glmmixedlasso}
Schelldorfer, J., L.~Meier, and P.~B{\"u}hlmann (2012).
\newblock {\em Generalized Linear Mixed Models with {Lasso}}.
\newblock https://r-forge.r-project.org/R/{?}group{\_}id=984.

\bibitem[\protect\citeauthoryear{Schelldorfer, Meier, and
  B{\"u}hlmann}{Schelldorfer et~al.}{2014}]{SchelldorferMeierBuhlmann2014}
Schelldorfer, J., L.~Meier, and P.~B{\"u}hlmann (2014).
\newblock {GLMMLasso}: An algorithm for high-dimensional generalized linear
  mixed models using {$L_1$}-penalization.
\newblock {\em Journal of Computational and Graphical Statistics\/}~{\em 23},
  460--477.

\bibitem[\protect\citeauthoryear{Seukep, Kolivras, Hong, Li, Prisley, Campbell,
  Gaines, and Dymond}{Seukep et~al.}{2015}]{Seukepetal2015}
Seukep, S.~E., K.~N. Kolivras, Y.~Hong, J.~Li, S.~P. Prisley, J.~B. Campbell,
  D.~N. Gaines, and R.~L. Dymond (2015).
\newblock An examination of the demographic and environmental variables
  correlated with {Lyme} disease emergence in {Virginia}.
\newblock {\em EcoHealth\/}~{\em 12}, 634--644.

\bibitem[\protect\citeauthoryear{Tibshirani}{Tibshirani}{1996}]{Tibshirani1996}
Tibshirani, R. (1996).
\newblock Regression shrinkage and selection via the {Lasso}.
\newblock {\em Journal of the Royal Statistical Society, Series B\/}~{\em 58},
  267--288.

\bibitem[\protect\citeauthoryear{Tseng and Yun}{Tseng and
  Yun}{2009}]{TsengYun2009}
Tseng, P. and S.~Yun (2009).
\newblock A coordinate gradient descent method for nonsmooth separable
  minimization.
\newblock {\em Mathematical Programming\/}~{\em 117}, 387--423.

\bibitem[\protect\citeauthoryear{Venables and Ripley}{Venables and
  Ripley}{2002}]{MASS}
Venables, W.~N. and B.~D. Ripley (2002).
\newblock {\em Modern Applied Statistics with S\/} (Fourth ed.).
\newblock New York: Springer.

\bibitem[\protect\citeauthoryear{{Virginia Department of Health}}{{Virginia
  Department of Health}}{2011}]{VDH2006-2011}
{Virginia Department of Health} (2011).
\newblock Reportable disease surveillance in {Virginia} (2006-2011 annual
  reports). {V}irginia {D}epartment of {H}ealth, {O}ffice of {E}pidemiology,
  {Richmond}, {VA}.
\newblock
  \url{http://www.vdh.virginia.gov/surveillance-and-investigation/virginia-reportable-disease-surveillance-data/}.
\newblock Accessed: 2018-09-10.

\bibitem[\protect\citeauthoryear{Yang}{Yang}{2007}]{Yang2007}
Yang, H. (2007).
\newblock {\em Variable selection procedures for generalized linear mixed
  models in longitudinal data analysis}.
\newblock {PhD} thesis, North Carolina State University, Raleigh, NC.

\bibitem[\protect\citeauthoryear{Yang and Zou}{Yang and
  Zou}{2012}]{YangZou2013}
Yang, Y. and H.~Zou (2012).
\newblock An efficient algorithm for computing the {HHSVM} and its
  generalizations.
\newblock {\em Journal of Computational and Graphical Statistics\/}~{\em 22},
  396--415.

\bibitem[\protect\citeauthoryear{Zhang}{Zhang}{2002}]{Zhang2002}
Zhang, H. (2002).
\newblock On estimation and prediction for spatial generalized linear mixed
  models.
\newblock {\em Biometrics\/}~{\em 58}, 129--136.

\bibitem[\protect\citeauthoryear{Zou}{Zou}{2006}]{Zou2006}
Zou, H. (2006).
\newblock The adaptive {Lasso} and its oracle properties.
\newblock {\em Journal of the American Statistical Association\/}~{\em 101},
  1418--1429.

\bibitem[\protect\citeauthoryear{Zou and Hastie}{Zou and
  Hastie}{2005}]{ZouHastie2005}
Zou, H. and T.~Hastie (2005).
\newblock Regularization and variable selection via the elastic net.
\newblock {\em Journal of the Royal Statistical Society, Series B\/}~{\em 67},
  301--320.

\bibitem[\protect\citeauthoryear{Zou and Zhang}{Zou and
  Zhang}{2009}]{ZouZhang2009}
Zou, H. and H.~H. Zhang (2009).
\newblock On the adaptive elastic-net with a diverging number of parameters.
\newblock {\em Annals of Statistics\/}~{\em 37}, 1733--1751.

\end{thebibliography}

\end{document}